\documentclass[letterpaper]{JHEP3}

\let\ifpdf\relax
\usepackage{ifpdf}

\usepackage[pdftex]{graphicx}
\usepackage{afterpage}

\usepackage{latexsym}
\usepackage{amssymb,amsmath}

\newcommand{\eeql}[1]{\label{#1}\eeq}
\newcommand{\refl}[1]{(\ref{#1})}

\newcommand{\uprm}{\ensuremath{U(1)'}}
\newcommand{\x}{\ensuremath{\times}}                                                
\newcommand{\ord}[1]{\ensuremath{\mathcal{O}(#1)}}
\newcommand{\veva}[1]{\ensuremath{\langle#1\rangle}}
\newcommand{\diag}{\mbox{diag}}

\def\simgt{\mathrel{\lower2.5pt\vbox{\lineskip=0pt\baselineskip=0pt
           \hbox{$>$}\hbox{$\sim$}}}}
\def\simlt{\mathrel{\lower2.5pt\vbox{\lineskip=0pt\baselineskip=0pt
           \hbox{$<$}\hbox{$\sim$}}}}

\title{Light Sterile Neutrinos and Short Baseline Neutrino Oscillation Anomalies}

\author{JiJi Fan\\
        Department of Physics, Princeton University, Princeton, NJ, 08540\\
        E-mail: \email{jijifan@princeton.edu}}
 \author{Paul Langacker \\
Department of Physics, Princeton University, Princeton, NJ 08540 \\
School of Natural Sciences, Institute for Advanced Study, Princeton, NJ 08540 \\
E-mail: \email{pgl@ias.edu}}

\abstract{We study two possible explanations for short baseline neutrino oscillation anomalies, such as the LSND and MiniBooNE anti-neutrino data, and for the reactor anomaly. The first scenario is the mini-seesaw mechanism with two eV-scale sterile neutrinos. We present both analytic formulas and numerical results showing that this scenario could account for the short baseline  and reactor anomalies and is consistent with the observed masses and mixings of the three active neutrinos. We also show that this scenario could arise naturally  from an effective theory containing a TeV-scale VEV, which could be related to other TeV-scale physics. 
The minimal version of the mini-seesaw relates the active-sterile mixings to five real parameters and favors an inverted hierarchy. It has the interesting property that the effective Majorana mass for neutrinoless double beta decay vanishes, while the effective masses
relevant to tritium beta decay and to cosmology are respectively around 0.2 and 2.4 eV.
The second scenario contains only one eV-scale sterile neutrino but with an effective non-unitary mixing matrix between the light sterile and active neutrinos. We find that though this may explain the anomalies, if the non-unitarity originates from a heavy sterile neutrino with a large (fine-tuned) mixing angle, this scenario is highly constrained by cosmological and laboratory observations.}

\newcommand{\beq}{\begin{equation}}
\newcommand{\eeq}{\end{equation}}
\newcommand{\beqs}{\begin{eqnarray}}
\newcommand{\eeqs}{\end{eqnarray}}
\newcommand{\gsim}{\stackrel{>}{_\sim}}

\newcommand{\calo}{{\cal{O}}}

\begin{document}
\section{Review of the Neutrino Oscillation Anomalies}
\label{anomalies}
Neutrino oscillation experiments with solar, atmospheric, reactor, and accelerator neutrinos have established a standard picture of three-flavor neutrino mixing and masses~\cite{Nakamura:2010zzi}:
\beqs
7.05 \times 10^{-5} \, {\rm eV}^2  \le & \Delta m_{21}^2& \le 8.34 \times 10^{-5} \, {\rm eV}^2, \nonumber \\
2.07 \times 10^{-3} \, {\rm eV}^2  \le& |\Delta m_{31}^2|& \le 2.75 \times 10^{-3} \, {\rm eV}^2, \nonumber \\
0.25 \le \sin^2 \theta_{12} \le 0.37,  \quad 0.36 \le &\sin^2 \theta_{23}& \le 0.67, \quad \sin^2 \theta_{13} <0.056 \quad {\rm at \, \,90\% \, \,C.L.} \nonumber 
\eeqs
However, over the past ten years, several anomalies have been observed in the short baseline experiments, suggesting deviations from the standard three-flavor picture. They are 
\begin{itemize}
\item{LSND~\cite{Aguilar:2001ty}: the search for $\bar{\nu}_\mu \to \bar{\nu}_e$ conducted using the Liquid Scintillator Neutrino Detector with $L/E$(m/MeV) = 0.4 - 1.5 (m/MeV), where $L$ is the distance between the neutrino source and detector and $E$ is the neutrino energy. A $3.8 \sigma$ excess of $\bar{\nu}_e$ candidate events was reported. The $\bar{\nu}_\mu$ arises from $\pi^+/\mu^+$ decay at rest: $\pi^+ \to \mu^+ \nu_\mu$ and $\mu^+ \to e^+ \nu_e \bar{\nu}_\mu$. The signal is defined as a positron with a 52 MeV endpoint from $\bar{\nu}_e p \to e^+ n$ with a correlated 2.2 MeV photon from neutron capture on a free proton $np \to d \gamma$. Another experiment, KARMEN~\cite{Armbruster:2002mp}, also searched for $\bar{\nu}_\mu \to \bar{\nu}_e$ using $\bar{\nu}_\mu$ from $\mu^+$ decaying at rest,
with $L/E$(m/MeV) $\sim$ 0.4 - 1.8 (m/MeV). The result is consistent with zero oscillation events and leads to a 90\% C.L. upper limit on the oscillation probability $P_{\bar{\nu}_\mu\to\bar{\nu}_e} < 6.5 \times 10^{-4}$. There is a small region of parameters consistent with both LSND and KARMEN~\cite{Church:2002tc}. }

\item{MiniBooNE~\cite{AguilarArevalo:2008rc, AguilarArevalo:2010wv, Djurcic:2012jf}: searches for both $\nu_\mu \to \nu_e$ and $\bar{\nu}_\mu \to \bar{\nu}_e$ were carried out with $L/E$ = 0.2 - 2.6 (m/MeV). In the neutrino search~\cite{AguilarArevalo:2008rc}, no excess of $\nu_e$ was found above 475 MeV\footnote{A 3$\sigma$ excess of electron-like events at lower energy, apparently unrelated to the LSND signal, was reported. The origin of the excess is unknown.}.  In the anti-neutrino search reported in 2010~\cite{AguilarArevalo:2010wv}, an excess of $\bar{\nu}_e$ events above 475 MeV was observed with 2.75$\sigma$. The 2011 updated result represents an increase in statistics of 52\% and still favors LSND-like $\bar{\nu}_\mu \to \bar{\nu}_e$ oscillations over a background only hypothesis at 91.1\% C.L. in the range $L/E$ = 0.2 - 1.13 (m/MeV)~\cite{Djurcic:2012jf}.}

\item{Reactor anomaly~\cite{Mention:2011rk}: A re-evaluation of the expected mean reactor anti-neutrino flux yields a prediction 3\% higher than what was previously assumed. This implies that all reactor neutrino experiments with $L < 100$m have observed a deficit of $\bar{\nu}_e$ events compared to the theory prediction, of 5.7\% on average, at 98.6\% C.L.}
\end{itemize}

These three experiments could potentially be interpreted as oscillations between eV-scale sterile neutrinos $\nu_s$ and the active neutrinos\footnote{Alternative explanations are summarized in~\cite{Akhmedov:2010vy}.}. While the  3+1 scheme (with only one eV-scale  $\nu_s$) could still not be consistent with all data, Kopp, Maltoni and Schwetz (KMS)~\cite{Kopp:2011qd} and independently Giunti and Laveder (GL)~\cite{Giunti:2011gz}, showed that with the new reactor flux evaluation, the global fit of these and other short baseline experiments improves greatly in a 3+2 scheme with two eV-scale sterile neutrinos. This is because the three active neutrinos are essentially degenerate on the relevant eV-scale, requiring two sterile neutrinos for a CP-violating difference between $\nu_\mu\rightarrow \nu_e$ and $\bar\nu_\mu\rightarrow\bar \nu_e$ oscillations.
As already pointed out in~\cite{deGouvea:2005er}, the 3+2 scheme could be accommodated in a mini-seesaw paradigm, which we will prove in more detail in this paper. Moreover, we will show that the mini-seesaw paradigm could  easily be realized in an effective theory involving a scalar with a TeV-scale vacuum expectation value (VEV). The most serious constraints on this scheme are from cosmological observations. Current data favors one or two light sterile neutrinos, but their masses are strongly constrained: one eV-scale sterile neutrino may be barely allowed, but  two are inconsistent with all observations unless certain cosmological assumptions are relaxed~\cite{Hamann:2010bk}.

Another possible explanation for the short baseline anomalies is a model in between the 3+1 and 3+2 schemes, in which one sterile neutrino is light and the other  is heavy~\cite{Nelson:2010hz}. In this case, only one  oscillation length is comparable to the range of the experiments.
The heavy neutrino is either not kinematically produced or its much shorter oscillation length is averaged over. The situation could be described by a non-unitary mixing matrix~\cite{Langacker:1988up,Antusch:2006vwa}
for the active neutrinos and one light sterile neutrino. Without the unitarity constraint, CP violation could be present even for two-neutrino oscillations~\cite{FernandezMartinez:2007ms}. Naively one might expect cosmology to be more friendly to such a scheme as only one eV-scale  sterile neutrino is present. However, we found that  the heavy sterile neutrino still needs to have a relatively large (fine-tuned) mixing angle with the active ones to generate large enough CP violation if one wants to explain all the data. Thus, this scheme is also highly constrained by cosmological/astrophysical observations as well as by laboratory experiments such as meson decays. 

Finally we  mention another anomaly that receives relatively less attention, the ``Gallium anomaly". The SAGE~\cite{Abdurashitov:2009tn} and GALLEX~\cite{Kaether:2010ag} experiments   used megacurie sources of $^{51}$Cr and $^{37}$Ar to calibrate $\nu_e + ^{71}\!\text{Ga} \to ^{71}\!\!\text{Ge} + e^-$. The data from these two experiments are consistent with each other and show a deficit compared to the theory predictions, which can be interpreted as a 2.7$\sigma$ indication of $\nu_e \to \nu_s$, where $\nu_s$ is a sterile neutrino with $\Delta m^2\sim 2.24$ eV$^2$~\cite{Giunti:2010zu}. If true, this has a subtle implication for the estimate of $\nu_e$ background in the MiniBooNE neutrino oscillation channel. However, as shown by~\cite{Conrad:2011ce}, the KARMEN and LSND cross section measurements for $\nu_e + ^{12}\!\text{C} \to ^{12}\! \text{N}_{gs}+e^-$ exclude the best fit point for the Gallium anomaly at 3.6$\sigma$. Thus, we do not put this anomaly on our motivation list.

The paper is organized as follows. In Section~2, we  first review the basic formalism for neutrino oscillations involving sterile neutrinos and the global best fits from the literature. Then we  examine the mini-seesaw mechanism and propose an effective theory that realizes the mini-seesaw naturally. A minimal version of the mini-seesaw, in which the active neutrino masses are due entirely to mixing, is especially predictive and is examined in detail. It strongly favors an inverted hierarchy (IH) but implies a vanishing effective neutrinoless double beta decay mass $m_{\beta\beta}$.  It is shown that this minimal version can accomodate the observed active neutrino mass differences and mixings and can successfully fit the LSND and MiniBooNE data. 
We conclude the section by discussing the cosmological constraints and possible ways to evade them. In Section~3, we discuss another possible scenario that may explain the short baseline anomalies with one light and one heavy sterile neutrino. We comment on the rather stringent cosmological and astrophysical constraints on this scenario and possible evasions. We conclude in Section~4.
Further details of the minimal mini-seesaw fit are given in the Appendix.

\section{The Mini-Seesaw Mechanism}
\subsection{Global Fits}
We  first review the basic formalism for neutrino oscillations. Neglecting the masses in the active neutrino sector and assuming that only two sterile neutrinos participate in the oscillation, the probability of electron neutrino appearance in a muon neutrino beam is 
\beqs
P_{\nu_\mu \to \nu_e} &=& 4 |U_{e4}|^2 |U_{\mu4}|^2\sin^2 x_{4}+4 |U_{e5}|^2 |U_{\mu5}|^2\sin^2 x_{5} \nonumber \\
&&+8 |U_{e4}| |U_{\mu4}| |U_{e5}| |U_{\mu5}|\sin x_{4}\sin x_{5}\cos(x_{5}-x_{4}-\delta),
\label{eq:oscillation}
\eeqs
where $U_{\alpha i}$ are the mixing matrix elements between active neutrinos $\nu_\alpha\, (\alpha = e, \mu)$ and steriles $\nu_i\,  (i=4,5)$, and the phases $x_{i}$ are 
\beq
x_{i} \equiv 1.27\left(  \frac{m_i^2}{{\rm eV^2}} \right)\left( \frac{L/E}{\rm{m/MeV}} \right).
\eeq
In propagation through matter the phases may also be altered by forward scattering due to the weak interactions and thus are also sensitive to $E$. The matter effects are generically small in short baseline experiments unless exotic forces exist, which we will neglect in this paper. 
The physical CP-violating angle is defined as
\beq
\delta \equiv \arg \left(\frac{U_{e5}U^*_{\mu5}}{U_{e4}U^*_{\mu4}}\right).
\eeq
For anti-neutrino oscillations, one only needs to replace $\delta$ by $-\delta$ in Eq.~\ref{eq:oscillation}.

As shown by~\cite{Kopp:2011qd} and \cite{Giunti:2011gz}, with the new reactor flux prediction, the global fit to all oscillation data (based on 2010 MiniBooNE anti-neutrino data) improves considerably when the existence of two light sterile neutrinos is assumed. The best fit points, given in Table 1, prefer two steriles with masses of order 1 eV; the active-sterile mixing angles of order 0.1;  and a non-zero CP violating phase $\delta$ in the sterile sector. 
\begin{table}[h]
\begin{center}
\begin{tabular}{|c|c|c|c|c|c|c|c|c|}
\hline
& $\Delta m^2_{41}$&$\Delta m^2_{51}$ &$|U_{e4}|$&$|U_{e5}|$&$|U_{\mu4}|$&$|U_{\mu5}|$&$\delta/\pi$ &$\chi^2$/dof\\
\hline
KMS& 0.47 & 0.87 & 0.128 & 0.138 & 0.165 & 0.148 & 1.64&110.1/130  \\
GL & 0.90 & 1.60 &0.13 &0.13 &0.13 &0.08 & 1.52&22.2/5 \\
\hline
\end{tabular}
\caption{Best global fit points of KMS~\cite{Kopp:2011qd} and of GL~\cite{Giunti:2011gz}.
(The GL result is obtained from parameter goodness of fit where the number of degrees of freedom corresponds to the number of parameters in common to the data sets. For more details, see~\cite{Maltoni:2003cu}.)}
\end{center}
\end{table}

\subsection{Active-Sterile Neutrino Mixing}
Most  extensions of the standard model (SM) which allow nonzero neutrino mass\footnote{For general reviews, see, e.g.,~\cite{GonzalezGarcia:2002dz, Mohapatra:2005wg,GonzalezGarcia:2007ib,Langacker:1226768}.} involve sterile (i.e., $SU(2)$-singlet) neutrinos. Consider, for example, the case of three left-chiral active neutrinos $\nu^0_L$ and $n$ left-chiral sterile antineutrinos
$N^{0c}_L$, and their respective  CP conjugates $\nu^{0c}_R$ and $N^0_R$,
where the superscript $^0$ indicates weak eigenstate. In general, these will have
a mass term
\beq
- {\cal L} = \frac{1}{2} \left (\bar{\nu}^0_L \,\, \bar{N}_L^{0c}\right)
\left(
\begin{array}{cc}
M_T & M_D \\
M_D^T & M_S \end{array}
\right)
\left(
\begin{array}{c}
\nu_R^{0c} \\
N^0_R\end{array}\right),\eeql{massterm}
where $M_T=M_T^T$ and $M_S=M_S^T$ are $3\x 3$ and $n \x n$-dimensional symmetric
Majorana mass matrices, while $M_D$ is a $3 \x n$-dimensional  Dirac mass matrix.
$M_T, M_D,$ and $M_S$ violate weak isospin by $1, 1/2,$ and $0$ units, respectively.
In the pure Majorana ($M_D=0$) and  pure Dirac ($M_T=M_S=0$) cases there is no mixing between
$\nu^0_L$ and $N^{0c}_L$ (or between $\nu^{0c}_R$ and $N^0_R$). In the ordinary Type I seesaw model (with the $M_S$ eigenvalues
much larger than $M_D$ and $M_T$, typically $\gtrsim \ord{\text{TeV}}$) there is mixing, but it is very small. Moreover, the heavy, predominantly sterile,
states decouple from the low energy theory. 
 Therefore, mixing between active and light sterile neutrinos of the same helicity, which could account for the anomalies described in Section \ref{anomalies}, requires the simultaneous presence of small Dirac mass terms {\em and} small Majorana mass terms\footnote{It could instead be due to two distinct kinds of small Dirac mass terms, such as  one which links active and sterile neutrinos, and another which
 links sterile left and right-chiral fields.} (see,  e.g.,~\cite{Langacker:1998ut}). We will therefore consider an effective low energy theory
 in which all of the entries in \refl{massterm} are very small (usually $\lesssim \ord{\text{eV}}$) after integrating out any heavy states. Especially attractive is the {\em mini-seesaw}, which is similar to the ordinary seesaw except that the eigenvalues of $M_S$ are in the eV range (and much larger than the elements of $M_D$ and $M_T$). This typically leads to relations between the active-sterile mixings and the mass eigenvalues consistent with the observations. For example, in the
 one-family version of \refl{massterm}, the $2\x 2$ mass matrix is
 \[ M=\left(
\begin{array}{cc}
m_T & m_D \\
m_D & m_S \end{array}
\right). \]
Taking $m_T=0$ for simplicity and $m_S \gg m_D$, the physical masses of the predominantly active and sterile states
are $m_1\sim m_D^2/m_S$ and $m_2 \sim m_S$, respectively, with an active-sterile mixing $\theta\sim m_D/m_S \sim (m_1/m_2)^{1/2}$. Thus, $m_D=\ord{0.1 \text{ eV}}$ and $m_S=\ord{ 1 \text{ eV}}$ implies $m_1= \ord{0.01 \text{ eV}}$
and $\theta=\ord{0.1}$.  The extension to several families and the roles of family mixing and $M_T$ are discussed below.

\subsection{Mini-Seesaw from Higher-Dimensional Operators}
 Barring fine-tuning, the small values of   $M_D$ and $M_{T,S}$ needed for active-sterile mixing or for the  mini-seesaw most likely imply that they are suppressed  by additional gauge, global, or discrete symmetries compared to the simplest expectations. The remaining small elements may usually be described\footnote{Another possibility is that the mass terms are exponentially suppressed by nonpertubative effects, such as D-brane instantons. For reviews, see~\cite{Blumenhagen:2009qh,Langacker:2011bi}.} by
 higher-dimensional operators (HDO) involving powers of $S/\Lambda$, where $S$ represents a SM singlet 
field (or fields) whose VEV breaks the symmetry. $\Lambda$ is the new physics scale,
typically involving heavy particles or string excitations that have been integrated out. The lowest-dimensional operators\footnote{The operator approach for neutrino masses has been developed generally in~\cite{Langacker:1998ut, Langacker:2011bi, Chen:2006hn,Sayre:2005yh}. Specific models that can lead to such operators include
string or \uprm\ motivated models~\cite{Langacker:1998ut,Langacker:2008yv}, mirror worlds~\cite{Foot:1995pa,Berezhiani:1995yi}, 
gauge-mediated supersymmetry breaking~\cite{Dvali:1998qy}, compositeness~\cite{ArkaniHamed:1998pf}, dynamical electroweak symmetry breaking~\cite{Appelquist:2002me}, warped extra dimensions~\cite{McDonald:2010jm}, and variations on conventional seesaw and flavor models~\cite{Ma:1995xk,Borzumati:2000fe,Babu:2004mj,Zhang:2011vh,Barry:2011wb}. } yielding a realistic spectrum are (in superpotential notation)
\beq
W_T  \sim  \frac{ (H_u L)^2}{\Lambda} ,\qquad
W_D  \sim  \frac{S}{\Lambda} H_u L N ,\qquad
W_S  \sim 
 \frac{ S^2}{\Lambda} N^2.
\eeql{operators} 
In the  mini-seesaw limit this implies
\[ \left(
\begin{array}{cc}
M_T & M_D \vspace*{4pt}\\
M_D^T & M_S \end{array}
\right) 
= \left(
\begin{array}{cc}
\calo(\frac{\nu^2}{\Lambda}) & \calo(\frac{S \nu}{\Lambda})\vspace*{4pt}  \\
 \calo(\frac{S \nu}{\Lambda}) & \calo(\frac{S^2}{\Lambda})   \end{array}
\right)
= \left(
\begin{array}{cc}
\calo(0.01) & \calo(0.1) \vspace*{4pt} \\
 \calo(0.1) & \calo(1)   \end{array}
\right),\]
where $\nu \sim 246$ GeV is the electroweak scale and we have taken
 $\veva{S} = \ord{\text{TeV}}$ and   $\Lambda \sim 10^{15}$ GeV. 
 Note that if both $W_D$ and $W_S$ are allowed by a multiplicative symmetry then so is
 $W_T$, and that 
 in the mini-seesaw limit
the mixing-induced contribution to the light eigenvalues, $ M_D^2/M_S\sim \nu^2/\Lambda$, is generically comparable to that from $W_T$.

\subsection{Active-Sterile Mixing Parameters in the Mini-Seesaw}
\label{sec:miniseesaw fit}
The implications of the general mass terms in \refl{massterm} and its (mini-)seesaw limit have been discussed in detail in~\cite{Casas:2001sr,deGouvea:2005er,Smirnov:2006bu,Donini:2011jh,Blennow:2011vn,Xing:2011ur}.
Here, we recount the relevant features for our analysis. In the general case the symmetric mass matrix in  \refl{massterm} 
can be diagonalized to yield $3+n$ Majorana mass eigenstates 
\[
\nu_L = {\cal A}_L^{\nu \dagger}
\left(
\begin{array}{c}
\nu_L^0 \\
N_L^c\end{array}\right),\]
where ${\cal A}_L^\nu$ is a $(3+n)  \times (3+n)$ unitary matrix and $\nu_L$ is a $(3+n)$-component vector. The analogous transformation for the $R$ fields is ${\cal A}_R^\nu = {\cal A}_L^{\nu *}$. In the important special case
$M_T=0$ and $n<3$, there will be $3-n$ massless active neutrinos\footnote{There can be additional massless states for
singular $M_D$, but we will not consider that case.}. Thus, $n=1$ or $2$ sterile neutrinos and $M_T=0$ would
be candidates for describing the NH or the IH for the active neutrinos, respectively.

In the seesaw limit, where the  eigenvalues of $M_S$ are both large compared to all entries of $M_D$ and $M_T$, one has
\[
 {\cal A}_L^{\nu \dagger}= {\cal A}_R^{\nu T}=
\left(
\begin{array}{cc}
A_L^{\nu \dagger} & 0 \\
0 & A_L^{N \dagger} \end{array}
\right)B_L^{\nu \dagger},\]
where $A_L^{\nu \dagger}$ ($A_L^{N \dagger}$) are $3 \times 3\ (n \times n)$ unitary matrices and 
\[
B_L^{\nu \dagger}= 
\left(
\begin{array}{cc}
I & -M_DM_S^{-1} \\
M_S^{-1\dagger} M_D^\dagger & I \end{array}
\right).\]
One finds that 
\[
B_L^{\nu \dagger}\left(
\begin{array}{cc}
M_T & M_D \\
M_D^T & M_S \end{array}
\right) B_L^{\nu *}= 
\left(
\begin{array}{cc}
M_T -M_DM_S^{-1}M_D^T & 0\\
0&M_S\end{array}
\right).\]

Without loss of generality, we can choose a basis in which $M_S$ is diagonal, with real and positive elements. We will first consider the limit $M_T=0$, which we refer to as the minimal mini-seesaw.
The mass matrix for the three active neutrinos is then
\beq m_\nu = M_DM_S^{-1}M_D^T = D D^T,  \eeql{mass1}
where $D=M_D M_S^{-1/2}$ is a $(3\x n)$-dimensional complex matrix, and the extra minus sign has been absorbed in a redefinition of the weak eigenstate phases.
$m_\nu$ can also be expressed as
\beq m_\nu =  A_L^\nu m_d A_L^{\nu T} = L L^T,  \eeql{mass2}
where  $m_d$ is a diagonal matrix of the   three light neutrino mass eigenvalues $m_i$, which can be taken to be real and positive by appropriate choices of mass eigenstate phases,
and $L\equiv A_L^\nu m_d^{1/2}.$ It is convenient to view $L$ as 
$(3\x n)$-dimensional: for $n< 3$ one can simply remove the  $3-n$ columns of zeros corresponding to $m_i=0$,
while for $n>3$ one can add $n-3$ extra columns of zeros.
We work in a basis for which the charged lepton
mass matrix is diagonal, in which case $A_L^\nu$ is just the active neutrino (PMNS) matrix~\cite{Maki:1962mu, Pontecorvo:1967fh}
\beq A_L^\nu= 
\begin{pmatrix}
1&0&0\\
0&c_{23}&s_{23} \\
0&-s_{23}&c_{23} \\
\end{pmatrix}
\begin{pmatrix}
c_{13} & 0 & s_{13} e^{-i \rho} \\
0&1&0\\
 -s_{13} e^{i \rho} &0&c_{13} \\
\end{pmatrix}
\begin{pmatrix}
 c_{12}&s_{12}&0\\
 -s_{12} & c_{12}& 0\\0&0&1\\
\end{pmatrix}
\Phi ,
\eeql{PMNS}
where $c_{ij}\equiv \cos \theta_{ij}$ and $\ s_{ij} \equiv \sin \theta_{ij}$ 
are leptonic mixings, $\rho$ is the CP-violating phase,
and $\Phi=\diag (e^{i \alpha_1},\,e^{i \alpha_2},\, e^{i \alpha_2}  )$ is a diagonal phase matrix.
The Majorana phases $\alpha_i, i=1,2,3$ are not observable in ordinary 3-flavor oscillations, but they do affect the active-sterile mixing. The charged lepton phases can be chosen so that one of the $\alpha_i$ is zero. We will use this freedom to choose $\alpha_1=0$. Then, $\alpha_3$ is unobservable for $m_3=0$.
The observational data is consistent with the tri-bimaximal form $s_{12}=1/\sqrt{3},\, \theta_{13}=0,\, \theta_{23}=\pi/4$~\cite{Harrison:2002er, Ma:2004zv}, i.e.,
\beq A_L^\nu=\begin{pmatrix}\phantom{-} \sqrt{{2\over 3}}  &\phantom{-} \sqrt{{1\over 3}}  & \phantom{-}0
 \\-\sqrt{{1\over 6}}&\phantom{-}\sqrt{{1\over 3}} &\phantom{-} \sqrt{{1\over 2}} \\ \phantom{-}\sqrt{{1\over 6}}& -\sqrt{{1\over 3}}&\phantom{-} \sqrt{{1\over 2}}
\end{pmatrix}
\Phi \approx
 \left(
\begin{array}{ccc}
\phantom{-}0.82 & \phantom{-}0.58 & 0 \\
-0.41 &\phantom{-} 0.58 &0.71 \\
\phantom{-}0.41 &- 0.58 & 0.71 \end{array}
\right) 
\Phi,
\eeql{pmnsval}
except possibly for recent hints from T2K, MINOS, and Double CHOOZ for a nonzero $\theta_{13}$~\cite{Abe:2011sj, Adamson:2011qu, Abe:2011fz}. 
We will use this form in our numerical illustrations, but have verified that small
modifications are not important for our purposes.

The active-sterile neutrino mixing is described by the $3\x n$ matrix
\[  U= i M_D M_S^{-1} =i D  M_S^{-1/2}, \]
where the $i$ results from the phase redefinitions mentioned above. 
From \refl{mass1} and \refl{mass2}, the most general solution for $D$ in terms of $L$ is
\beq D=L R(z_k), \eeql{Dsoln}
where $R$ is an orthogonal $n\x n$ complex matrix, which depends on $n(n-1)/2$ complex parameters $z_k$  as well as signs. Thus, the mixings can be predicted in terms of the PMNS matrix, the light and heavy mass eigenvalues, and the  $z_k$. 

For $n=1$ sterile neutrino one has a NH, with  $m_1=m_2=0$, and with $m_3\sim 0.047$ eV determined from the atmospheric and long baseline neutrino oscillations. The active-sterile mixings are given by
\beq
 U_{\alpha 4} =i\frac{M_D^{\alpha 4}}{{M_4}}=\pm i A_L^{\nu \alpha 3}\sqrt{\frac{ {m_3}}{{M_4}}},  \quad \alpha=e, \mu, \tau.
\eeql{n1}
Because of the stringent upper limit  on $A_L^{\nu e 3}\lesssim 0.24$, $ U_{e 4} $
is too small to account for the LSND/MiniBooNE anomaly~\cite{deGouvea:2005er} (even ignoring the need for CP violation to differentiate 
$\nu_\mu$ from $\bar\nu_\mu$ and the lack of a Solar mass splitting between $m_1$ and $m_2$).

For $n>1$
\beq 
\boxed{ U_{\alpha i} =i \frac{M_D^{\alpha i}}{M_i}=i A_L^{\nu \alpha j}\sqrt{ m_j}\,  R_{ji}\, \frac{1}{\sqrt{M_i}}}, 
\eeql{ng1}
where $  \alpha=e, \mu, \tau$; $j=1, \cdots,n$; and $i= 4, \cdots, 3+n$.  In particular, for $n=2$ 
there is one massless neutrino. In principle one could have either a NH or IH for the three light neutrinos.
However, for the normal case it is difficult to obtain large enough active-sterile mixings because of the small values of 
 $A_L^{\nu e 3}$ and $m_2$. We therefore consider
 an IH  with $m_3=0$, with the active-sterile mixings determined by the
masses and PMNS matrix, as well as one complex parameter\footnote{The discrete sign in $R$
can be absorbed by redefining the sign of a mass eigenstate field. For similar reasons it suffices to
restrict $0 \le \alpha_2 < \pi$.} $z$, i.e.,
\beq
R(z)= \begin{pmatrix}
\cos z& \sin z\\
 -\sin z & \cos z
\end{pmatrix}.
\eeql{ne2}
Because of the possible Majorana phases in $A_L^\nu$ and because $R$ is complex, there is room for CP violation in the active-sterile mixing. Note that all values of the PMNS parameters and of  the mass eigenvalues (with $m_3=0$) are consistent with the minimal mini-seesaw parametrization.

To test whether Eq.~\ref{ng1} could account for the LSND and MiniBooNE results, we take the experimental data to construct a $\chi^2$ function
\beq
\chi^2(z, \alpha_2, M_4, M_5)=\sum_i\frac{\left(P_i^{\rm theory}(z, \alpha_2, M_4, M_5)-P_i^{\rm exp}\right)^2}{\sigma_i^2},
\label{eq:chisquare}
\eeq
where $P_i^{\rm exp}$ is the oscillation probability for bin $i$ from the experimental data and $\sigma_i$ is the corresponding error. $P_i^{\rm theory}(z, \alpha_2, M_4, M_5)$ are computed from Eq.~\ref{eq:oscillation}, \ref{ng1}, \ref{ne2}. We assume the IH, so  $\alpha_2$ is the only relevant unknown active neutrino parameter, and set the PMNS matrix to be Eq.~\ref{pmnsval}. We include 8 bins each for LSND, MiniBooNE neutrino and MiniBooNE anti-neutrino data. We do not use MiniBooNE data for $E <$ 475 MeV ($L/E > 1.13$ m/MeV) because of the unexplained excess in the low energy bins\footnote{Comparison between data with $E >$ 475 MeV and $E>$ 300 MeV are performed in~\cite{Maltoni:2007zf,Giunti:2011hn}.}. For the MiniBooNE anti-neutrino search, we performed two separate fits, using just the 2010 results or including the 2011 data. Before showing the results, we want to emphasize the differences between our fit and the global fits in~\cite{Kopp:2011qd,Giunti:2011gz}. Though our fit is crude in the sense of not including all of the oscillation data available, it is a more direct test of the minimal mini-seesaw mechanism as an explanation for the short baseline oscillation anomalies. The global fits employ all physical parameters: the sterile neutrino masses, mixing angles and CP-violating phase, while the parameters in the minimal mini-seesaw are more restricted.  Our fits use 5 real parameters characterizing the mini-seesaw mechanism: the complex angle $z$, active neutrino CP-violating phase $\alpha_2$, and two sterile neutrino masses. The mixing angles between the active and sterile neutrinos and the sterile CP-violating phase can be determined from these 5 parameters using  Eq.~\ref{ng1}. Even though we do not employ the full data set, we require the mixings between sterile and active neutrinos to be smaller than 0.15 in the numerical evaluation of the best fit to avoid conflicting with the reactor data or with  other neutrino oscillation experiments with null results.

\begin{figure}[!h]
 \begin{tabular}{cc}
\includegraphics[scale=0.55]{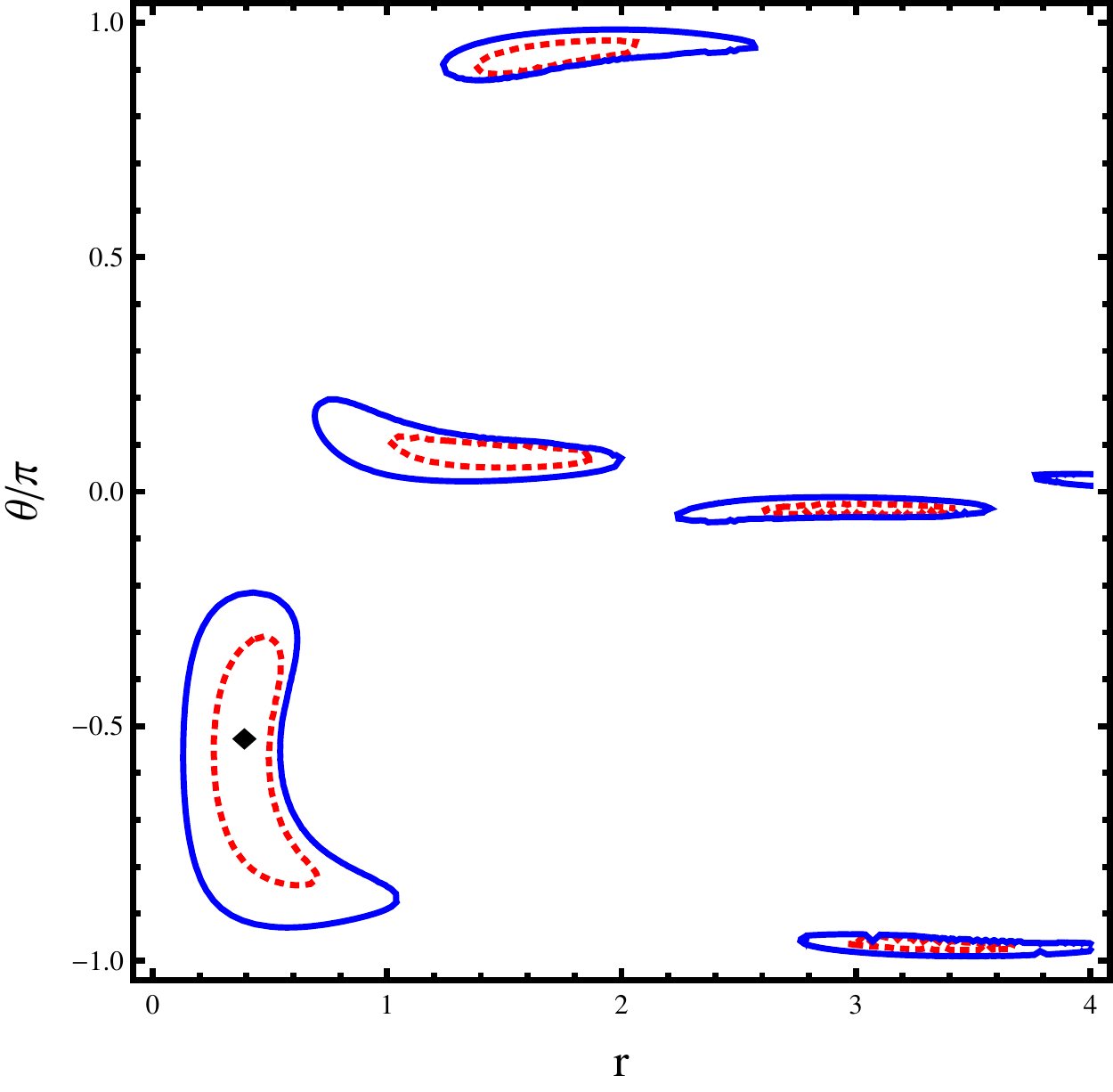} 
&\includegraphics[scale=0.55]{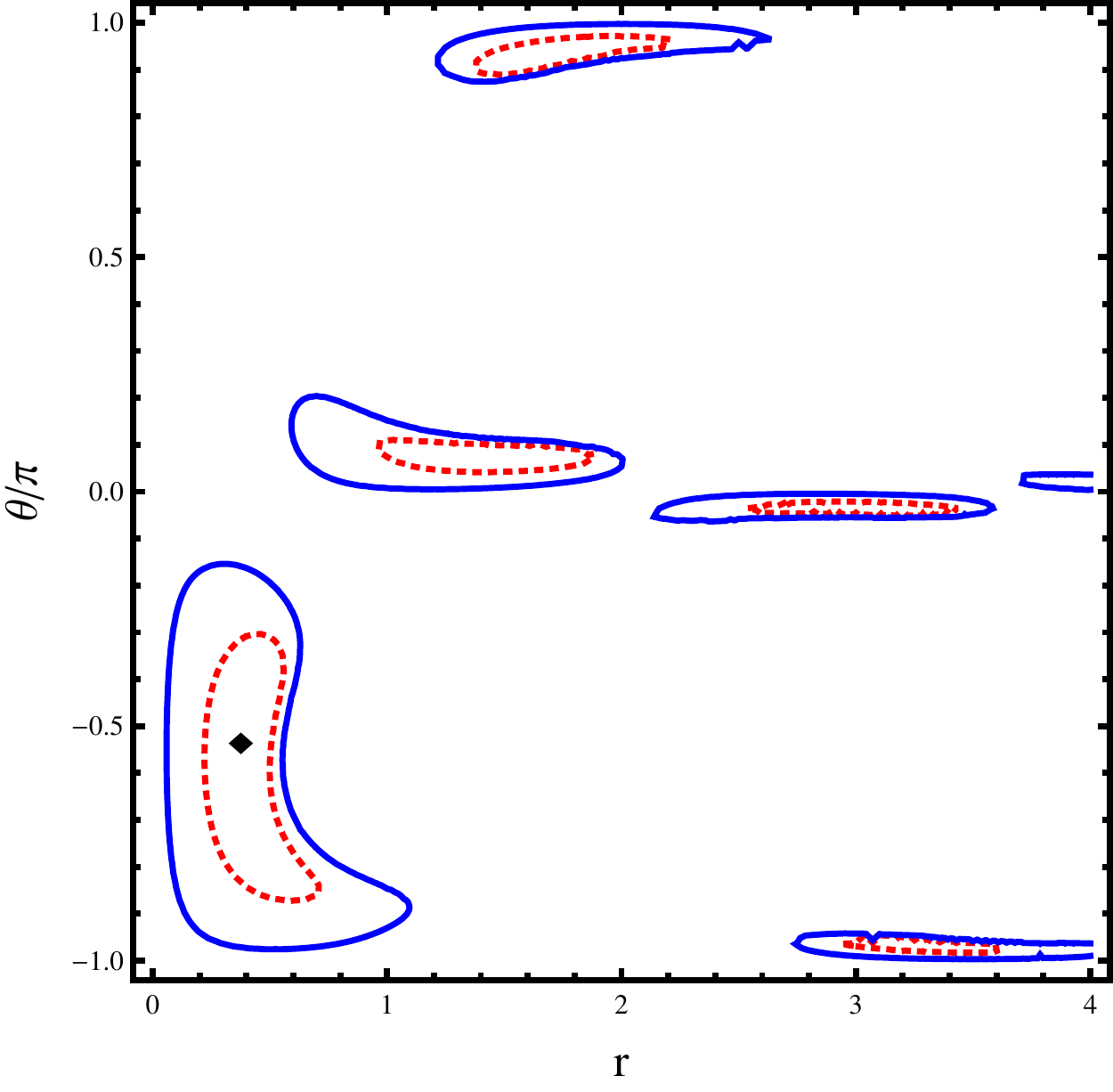} 
\end{tabular}
\caption{The allowed regions of $z=re^{i\theta}$ at  68\% (red, dotted) and 95\% (blue, solid) C.L., fixing $\alpha_2, M_4, M_5$ to  the best fit values. Left: fit using the 2010 MiniBooNE anti-neutrino data. Right: fit including 2011 data. They are almost identical. The best fit points are denoted by diamonds. }
\label{fig:ss_chisquare}
\end{figure}

The best fit results are presented in Table~\ref{fitresults}. They are in the right ballpark compared to the global fit values. In particular, the predicted parameters are quite close to the GL fit values.
We also show the allowed regions of $z=re^{i\theta}$ at 68\% and 95\% C.L. from our simple fits in Figure~\ref{fig:ss_chisquare}, fixing the other three parameters to  the best fit values. (The best fit constraint $|U_{\alpha i}| < 0.15$
is {\em not} enforced in these contours.) The two fits using 2010 and 2011 data give similar results, though the $\chi^2$ is larger for the 2011 data. The predicted transition probabilities as a function of $L/E$ for both LSND and MiniBooNE  at our best fit point are presented in Figure~\ref{fig:spectrum}. Further details are given in the Appendix.

\begin{table}[h]
\begin{center}
\begin{tabular}{|c|c|c|c|c|c|c|c|c|c|c|}
\hline
&$z$&$\alpha_2$&$\Delta m^2_{41}$& $\Delta m^2_{51}$&$|U_{e4}|$&$|U_{\mu4}|$ &$|U_{e5}|$&$|U_{\mu5}|$&$\delta/\pi$&$\chi^2$/dof \\
\hline
MMS(2010) & 0.39 $e^{-i 0.53\pi}$ &2.01& 0.89 & 1.78 & 0.15&0.15&0.07&0.15 & 1.25 &18.6/19 \\
MMS(2011) & 0.38 $e^{-i 0.54\pi}$& 1.92 & 0.89 & 1.76& 015 &0.15&0.07&0.15 & 1.21 &24.3/19\\
\hline
KMS& &  & 0.47 & 0.87& 0.128&0.165 & 0.138&0.148 & 1.64&110.1/130 \\
GL & & &0.90&1.60& 0.13&0.13&0.13&0.08&1.52&22.2/5 \\
\hline
\end{tabular}
\caption{Best fit points using the minimal mini-seesaw (MMS) formalism derived in the text. The first row uses the 2010 MiniBooNE anti-neutrino results while the second one includes the 2011 data.
For comparison, we also show the global fit results KMS~\cite{Kopp:2011qd} and GL~\cite{Giunti:2011gz}.} 
\label{fitresults}
\end{center}
\end{table}

 \begin{figure}[!h]
\begin{tabular}{cc}
\includegraphics[scale=0.43]{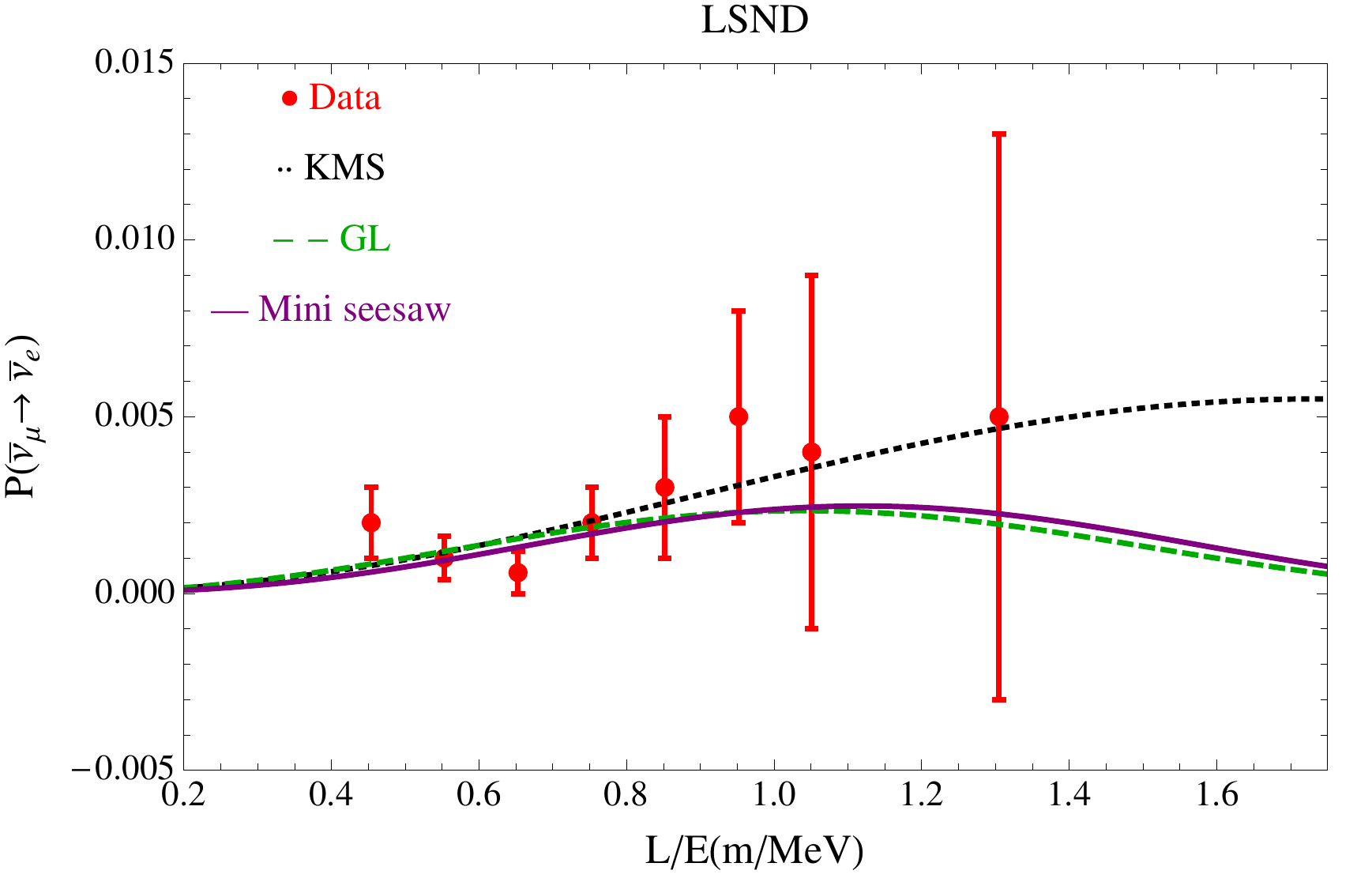}& 
\includegraphics[scale=0.43]{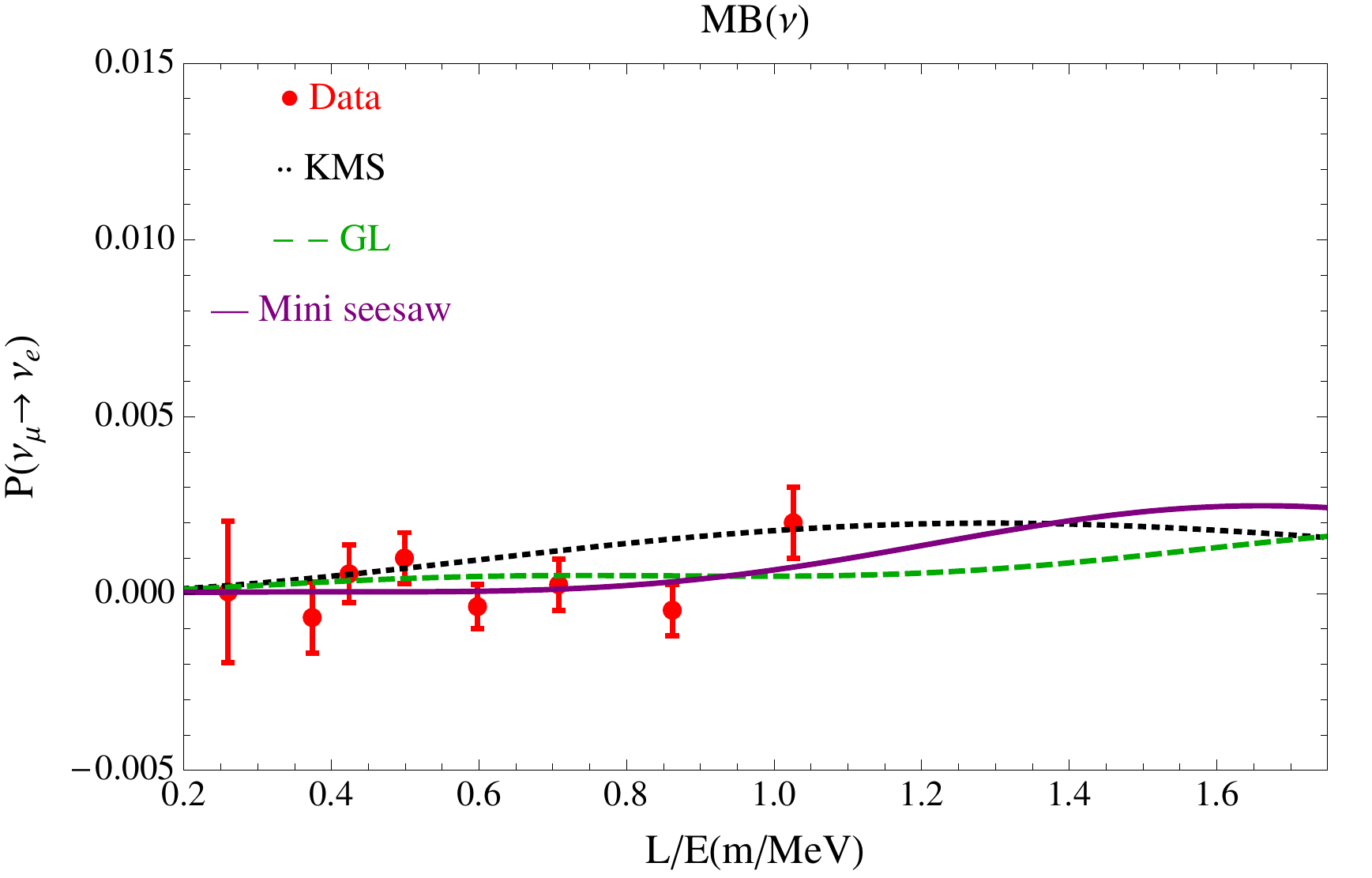} \\
\includegraphics[scale=0.43]{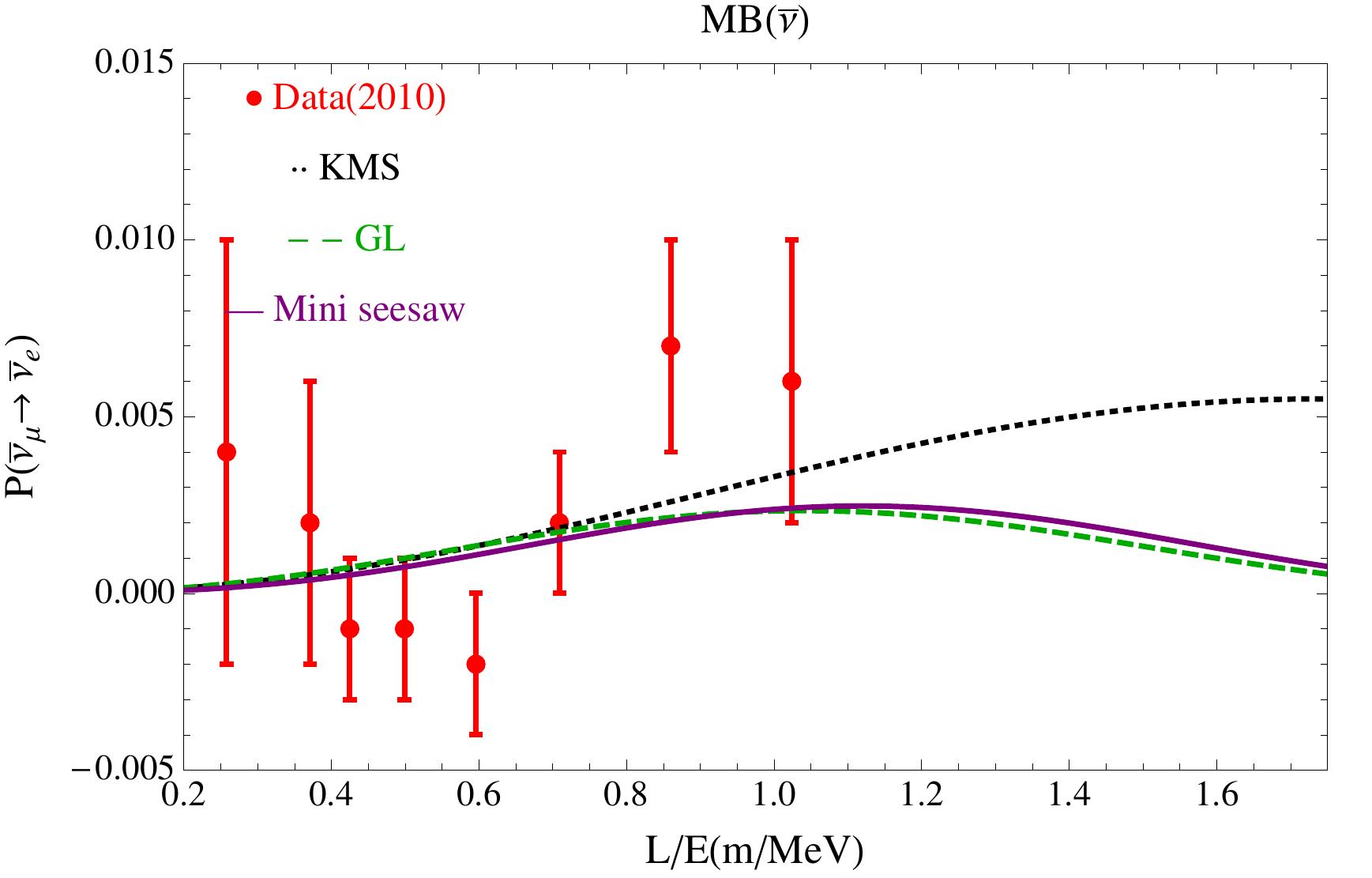} &
\includegraphics[scale=0.43]{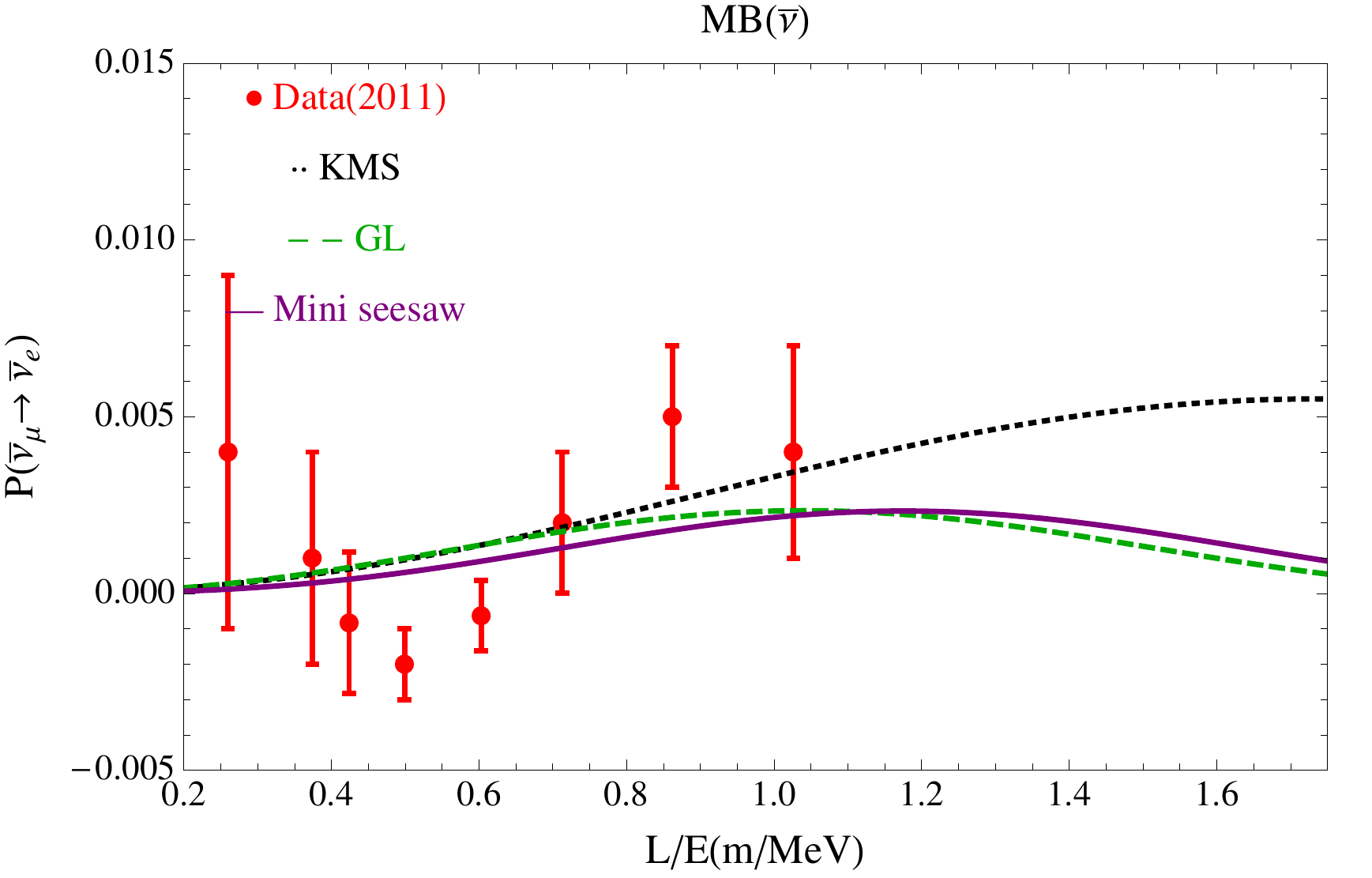}
\end{tabular}
\caption{Predicted transition probabilities for LSND, MiniBooNE neutrino and MiniBooNE anti-neutrino oscillations from the KMS~\cite{Kopp:2011qd} and GL~\cite{Giunti:2011gz}  global best fit points and from the best fit point  using the minimal mini-seesaw formula. }
\label{fig:spectrum}
\end{figure}

We have emphasized that one typically expects $M_T$ to be non-zero and, in the mini-seesaw limit, to yield contributions to the  active neutrino masses comparable to  those induced by the mixing. $M_T\ne 0$ of course reduces the predictabilty of the theory.  Assuming that all of the mass terms are of the order of magnitude suggested by \refl{operators} this  additional freedom  could allow a NH or degenerate scenario for the active neutrinos. For more general scenarios, e.g., with different $S$ fields or different dimensions of the effective operators for different families, one could  allow one or more sterile neutrinos considerably heavier than an eV. The mixing of the light active neutrinos with the heavier sterile states would typically be strongly suppressed, but relatively large mixing angles (so as to still be relevant to LSND/MiniBooNE) would  be possible by fine-tuning.

\subsection{Constraints and Implications}
\subsubsection{Laboratory Constraints}
In this section we comment on two laboratory constraints on neutrino masses from tritium $\beta$ decay, $^3 \text{H} \to ^3\!\text{He}\, e^-\bar{\nu}_e$, and neutrinoless double $\beta$ decay ($\beta\beta_{0\nu}$), i.e., $nn\to ppe^-e^-$. The shape of the electron spectrum near the endpoint in  tritium $\beta$ decay puts a stringent kinematic limit on the effective $\nu_e$ mass-squared 
\beq
m_{\beta}^2 \equiv \sum_i |{\cal A}_{ei}|^2 m_i^2,
\eeq
which is bounded to be in the range $m_{\beta}^2=(1.1\pm2.4)\,{\rm eV}^2$ from experiments in Mainz~\cite{Kraus:2004zw} and Troitsk~\cite{Lobashev:2001uu}. The Karlsruhe Katrin experiment should improve the sensitivity down to around $(0.2 \,{\rm eV})^2$ at the 90\% level~\cite{Thummler:2011zz}. The sterile neutrino state could have two effects on the electron spectrum. It could produce a kink in the spectrum of order $|U_{ei}|^2$ at energy $E_0-m_i$, where $E_0=18.6$ keV is the endpoint energy and $m_i$ is the sterile neutrino mass. As a result, it would be accompanied by a suppression of events at the endpoint of size $1-|U_{ei}|^2$. A detailed study has been performed in~\cite{deGouvea:2006gz}. While current experiments do not exclude an eV-scale sterile neutrino with mixing angle $\sim 0.1$, Katrin would start to be sensitive to this parameter region.  As shown in Table \ref{table: lightconstraints} the value of $m_\beta$ for our best fit point in the minimal mini-seesaw is 0.18 eV.

\begin{table}[h]
\begin{center}
\begin{tabular}{|c|c|c|c|}
\hline
& 3+2 (eV) & 3+2 MSS (eV) & EXP (eV) \\
\hline
$m_\beta$& $\sim 0.2$ &0.18   &$(1-2) \to 0.2$ \\
$m_{\beta\beta}$&$0-0.08 $& 0 & $(0.2 - 0.7) \to (0.01 - 0.03)$   \\
$\Sigma$& $\sim 2 $ & 2.4 &$(0.5- 1) \to (0.05-0.1)$ \\
\hline
\end{tabular}
\caption{Major constraints on the sterile neutrino masses. The second column lists typical
values for generic (non-degenerate) 3+2 schemes, while the third are the results at the best fit point of the
minimal mini-seesaw. $m_\beta$ and $\Sigma$ are dominated by the effects of the eV-scale sterile neutrinos.
The last column is the range probed by current and future experiments.}
\label{table: lightconstraints}
\end{center}
\end{table}

The process $\beta\beta_{0\nu}$ constrains the effective Majorana mass in the presence of mixing between light Majorana neutrinos,
\beq
m_{\beta\beta} \equiv \sum_i ({\cal A}_{ei})^2 m_i= (M_T)_{11},
\eeq
where the second equality is only true when all of the mass eigenstates are light compared to an MeV. In the particular minimal case we discussed above, $M_T=0$ and one expects
$m_{\beta\beta} =0$ due to a cancellation between the light and eV-scale states. This is in contrast to generic $3+2$ models~\cite{Barry:2011wb},
including the mini-seesaw with $M_T \ne 0$, in which  $m_{\beta\beta}$ can vary from 0 to around $0.08$ eV or $0.04$ eV for the IH or
the NH, respectively (both cases assuming a massless lightest state). The best current upper limit\footnote{One experiment~\cite{KlapdorKleingrothaus:2006ff} claims to observe a positive signal, corresponding to $m_{\beta\beta}\sim 0.3$ eV, but this has not been confirmed.} is from the Cuoricino experiment~\cite{Arnaboldi:2008ds} in the Gran Sassso Laboratory, which obtains $m_{\beta\beta} < (0.19 - 0.68)$ eV at 2$\sigma$, with the range due to the nuclear matrix element. 
Future experiments should be sensitive down to (0.01 - 0.03) eV~\cite{Bucci:2011zz, Gornea:2011zz, Aalseth:2011zz, Flack:2011zz, Meierhofer:2011zz, CANDLES:2011aa}, which start to constrain $M_T$.

Other existing and future implications of mixing with eV-scale sterile neutrinos are considered in ~\cite{deGouvea:2006gz,Razzaque:2011ab,Hernandez:2011rs,Gandhi:2011jg,Barger:2011rc,deGouvea:2011zz,Bhattacharya:2011ee}.

\subsubsection{Cosmological Constraints}
Though the 3 active + 2 eV-scale sterile neutrino model fits nicely into the mini-seesaw paradigm and could be explained easily by an effective theory containing a SM singlet with a TeV-scale VEV, it is nontrivial to accomodate two eV-scale sterile neutrinos into a consistent cosmological history. On one hand, current cosmic microwave background (CMB) and large-scale structure (LSS) observations show a slight preference for additional relativistic degrees of freedom beyond the SM prediction. The combination of the CMB, measurement of the Hubble parameter from HST observations and baryon acoustic oscillation (BAO) data shows that the effective number of relativistic degrees of freedom $N_{eff} $ is greater than the SM value $3.046$ at $98.4\%$ C.L.~\cite{Hou:2011ec}. By further combining with the Atacama Cosmology Telescope (ACT) power spectrum measurement, the number is estimated to be $N_{eff}=5.3\pm1.3$ (68\% C.L.)~\cite{Dunkley:2010ge}. Similar analyses that show preferences for light sterile neutrinos can be found in~\cite{GonzalezGarcia:2010un, Giusarma:2011zq}. However, it is argued in~\cite{GonzalezMorales:2011ty} that once the prior-dependence in the cosmological analysis is removed, the latest cosmological data show no evidence for deviations from the SM value of neutrino species. 
In the near future, data from the {\em Planck} satellite will be able to reduce the error in $N_{eff}$ to 0.20 and settle the issue~\cite{Bashinsky:2003tk}. 

On the other hand, the Wilkinson Microwave Anisotropy Probe (WMAP) 7-year data, small-scale CMB observations,  Sloan Digital Sky Survey (SDSS) 7th data and the present Hubble constant, set a 95\% C.L. limit on the sum of two light sterile neutrino masses~\cite{Hamann:2010bk}
\beq
\Sigma_S=\sum_{i=4}^5   |m_i|<0.9 \,{\rm eV} \quad (95\% \, {\rm C.L.}),
\eeq
which implies $\Sigma\equiv \sum_i  |m_i|\lesssim 1 $ eV for the non-degenerate IH, which is small compared to the typical values in Table~\ref{table: lightconstraints}.

It is also well known that if two eV-scale steriles are fully thermalized prior to neutrino decoupling, they would increase the Hubble rate and result in an overproduction of  $^4$He in the epoch of big bang nucleosynthesis (BBN). A combined analysis of $^4$He and deuterium data sets a 95\% C.L. limit on $N_s < 1.26$~\cite{Hamann:2011ge} with a best-fit value $N_s = 0.86$. So while one fully thermalized neutrino is slightly favored over none, two fully thermalized sterile neutrinos with  masses around 1 eV are not fully compatible with all the astrophysical data unless certain assumptions of these analysis are relaxed.

There are several (highly speculative/creative) loopholes to these cosmological constraints.
 One possibility is that there exists a common non-zero chemical potential for the active neutrinos $\xi$~\cite{Kang:1991xa, Barger:2003rt,Mangano:2010ei}. In the BBN epoch, a non-zero chemical potential would drive the reactions $n+\nu_e \leftrightarrow p+e^-,\  n+e^+ \leftrightarrow p+\bar{\nu_e}$ in one direction and change the ratio of neutrons to protons in thermal equilibrium to
\beq
\frac{n}{p} =\exp\left(-\frac{m_n-m_p}{T}-\xi\right).
\eeq
The additional suppression would compensate the effects of  an increase of the relativistic degrees of freedom. Assuming two light sterile neutrinos, a positive non-zero chemical potential is needed $0.03<\xi<0.14$ at 95\% C.L. with a best-fit value of $\xi=0.064$~\cite{Hamann:2011ge}. 
Such a large chemical potential would correspond to an enormous lepton asymmetry of the same order as $\xi$
($\sim 10^8\, \times$  the baryon asymmetry), but might be generated from a much smaller
initial asymmetry due to non-linear effects associated with the sterile neutrinos ~\cite{Foot:1995bm,Foot:1995qk, Abazajian:2004aj}.

Another possibility discussed in the literature is that the sterile neutrino has a time-dependent mass~\cite{Fardon:2003eh, Kaplan:2004dq, Bi:2003yr}. This could be achieved if the singlet that generates the sterile neutrino mass has a medium-dependent VEV, i.e., a VEV proportional to the number density $n_\nu$ of the active neutrinos: $\langle S \rangle \propto n_\nu$. Then $M_S^z=M_S(1+z)^n$  in our effective model,  where $M_S^z$ is the mass in the epoch with redshift $z$ and $n=6$ if the sterile Majorana mass is generated from a non-renormalizable operator $S^2N^2$~\footnote{The discussion here is slightly different from that in~\cite{Smirnov:2006bu}, in which $n=3$, as the sterile mass is assumed to be from a renormalizable operator $SN^2$. It is also assumed in~\cite{Smirnov:2006bu} that  the Dirac mass mixing the sterile and active neutrinos is generated by the Higgs VEV and does not change with time. Thus the induced contribution to the light mass $M_DM_S^{-1}M_D^T$ and the active-sterile mixing angle $M_D M_S^{-1}$ decrease at early  times. However in our effective theory, the Dirac mass also decreases with time $M_D \propto (1+z)^3$ and thus the induced contribution to the light neutrino mass is constant while the mixing angle still decreases with time.}. Then at the recombination epoch with $z \sim 1100$, the sterile neutrino mass becomes $10^9$ GeV for the present mass of 1 eV. Thus the production of sterile neutrinos are highly suppressed in the early Universe and their abundance is negligible. 

A third possibility is to have late time phase transitions to suppress the sterile neutrino masses and mixings until after they decouple at $T < 1$ MeV. This could be achieved in models with low-scale breaking of global symmetries~\cite{Chacko:2004cz}.

\section{Heavy Sterile Neutrino}
\subsection{The Allowed Parameter Region}
There is another interesting possibility~\cite{Nelson:2010hz}, that the LSND and MiniBooNE results could be consistent with the paradigm with one eV-scale sterile neutrino and another very heavy sterile with mass ranging from $\calo(10)$ eV to $\calo(10)$ GeV. In the presence of a coherent heavy sterile neutrino $\nu_5$, $x_5$ varies very rapidly and must be averaged over. Furthermore, the heavy neutrino could be produced incoherently and/or with a reduced phase space. If it is sufficiently heavy, above the energy of the experiment, it is not produced at all. Following~\cite{Nelson:2010hz}, we introduce two parameters: the mixing ratio $r$ and the CP odd quantity $\beta$,  defined as 
\beqs
r &\equiv &\frac{|U_{e5}U_{\mu5}^*+U_{e4}U_{\mu4}^*|}{|U_{e4}U_{\mu4}^*|} \nonumber \\
\beta  &\equiv & \frac{1}{2} \tan^{-1}\left(\frac{\sin \delta |U_{e5}U_{\mu5}|}{|U_{e4}U_{\mu4}|+\cos\delta|U_{e5}U_{\mu5}|}\right)\nonumber \\
\eeqs
so that 
\beq
re^{2i\beta}=\frac{U_{e5}U_{\mu5}^*+U_{e4}U_{\mu4}^*}{U_{e4}U_{\mu4}^*}.
\label{eq:mixing5}
\eeq
Also taking into account  the possibly reduced phase space by including another free parameter $a$, which varies from 0 (no production) to 1 (no suppression), we have the appearance probability 
\beqs
P_{\nu_\mu \to \nu_e} =  |U_{e4}|^2 |U_{\mu4}|^2 \left((1-r)^2+a\left((1-r)^2+4r\sin^2\beta\right)+4r\sin^2(x_4+\beta)\right),
\label{eq:oscillation2}
\eeqs
with $\beta \to -\beta$ for anti-neutrinos.

\begin{figure}[!h]
\begin{tabular}{ccc}
\includegraphics[scale=0.4]{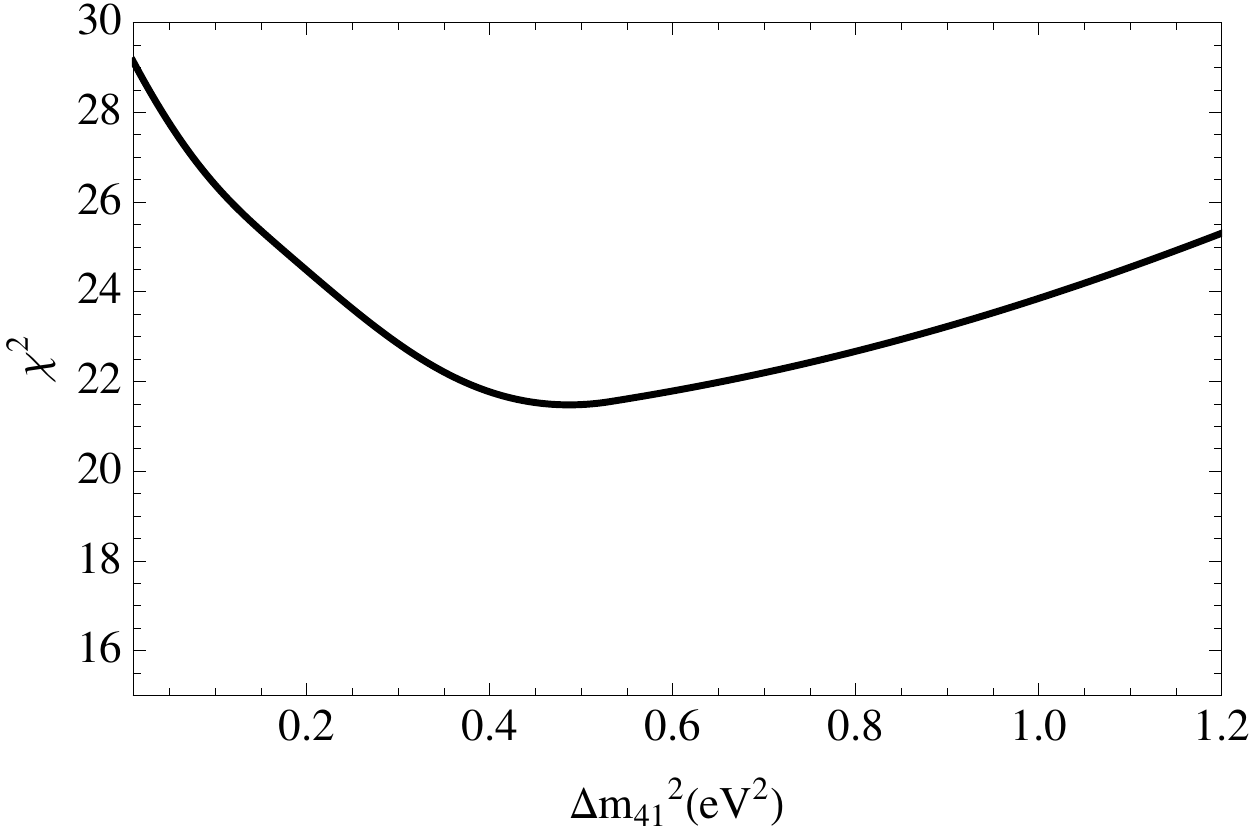}& 
\includegraphics[scale=0.4]{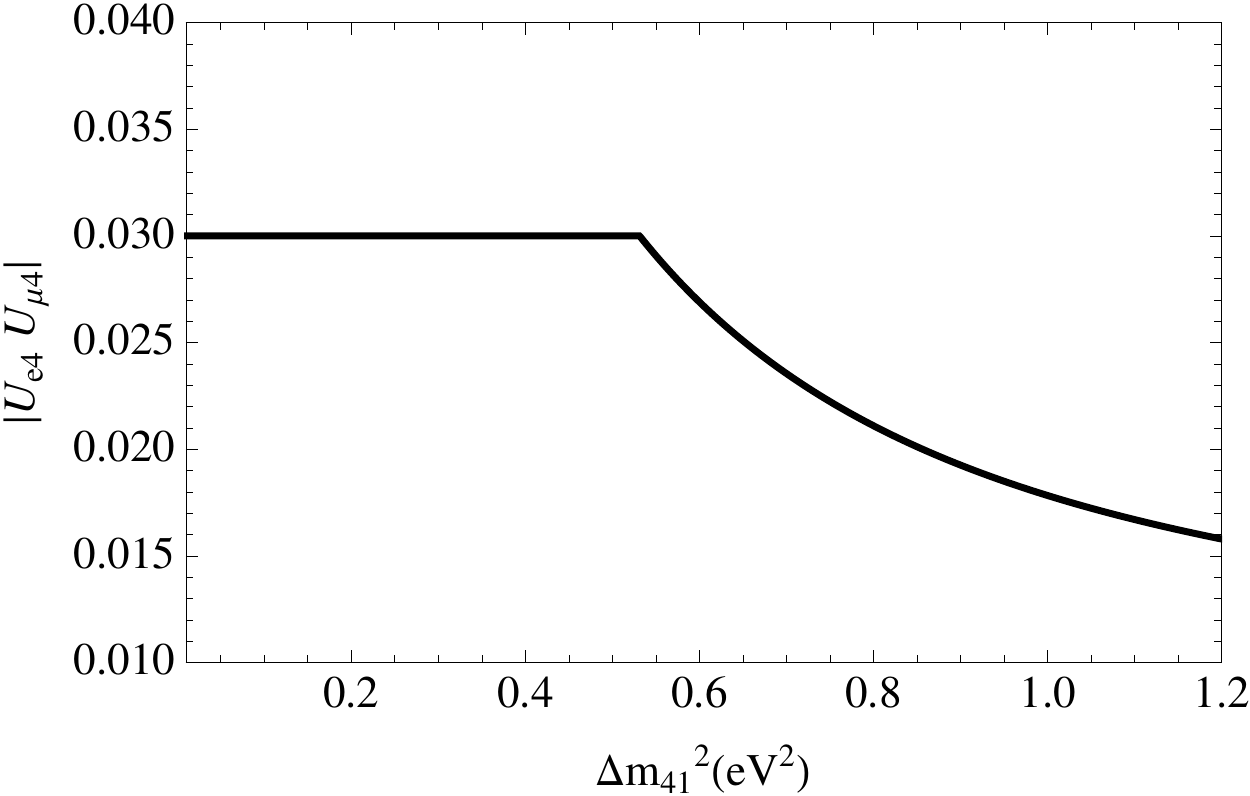} &
 \includegraphics[scale=0.4]{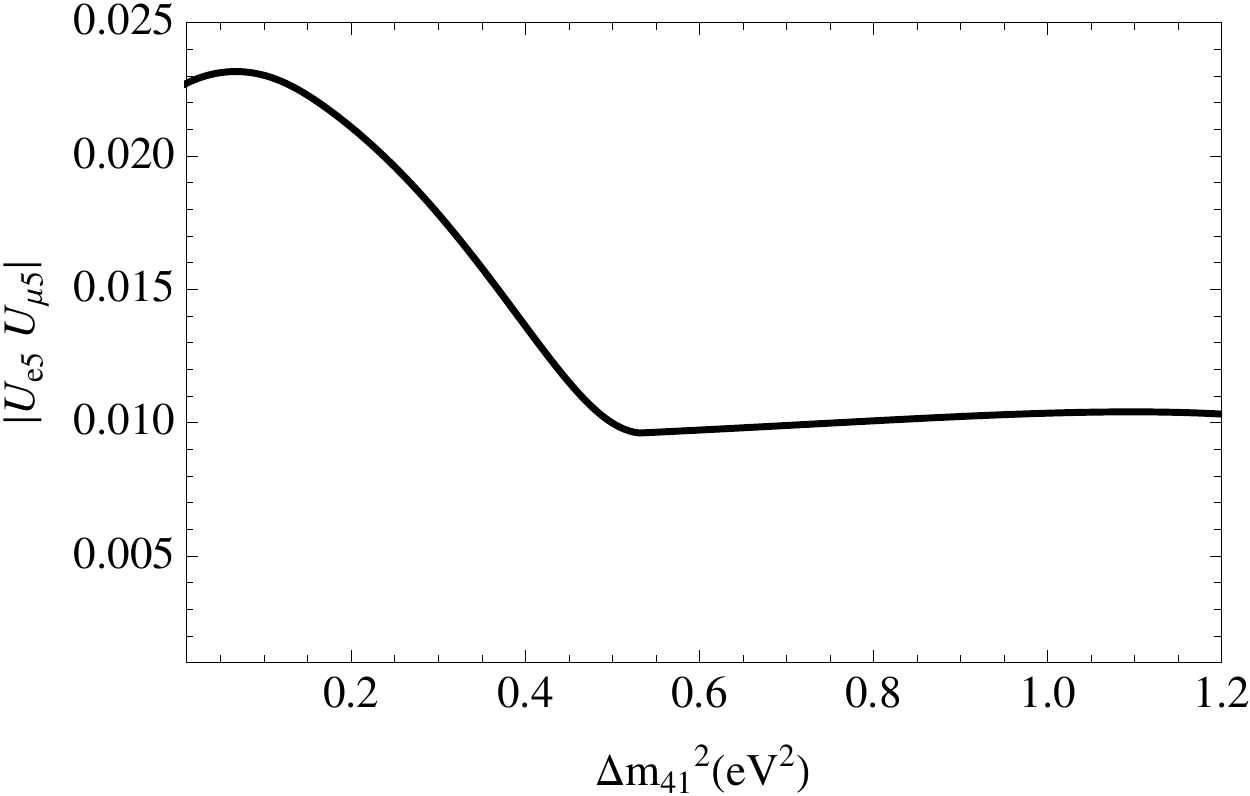}
\end{tabular}
\caption{The minimal values of $\chi^2$ and best fit values of $|U_{e4}U_{\mu4}|$ and $|U_{e5}U_{\mu5}|$ as a function of $\Delta m_{41}^2$ in the scenario with a light and a heavy sterile neutrino. The constant value of $|U_{e4}U_{\mu4}|$ for small $\Delta m_{41}^2$ is due to the imposed constraint $|U_{e4}U_{\mu4}| \leq 0.03$.}
\label{fig:chisquare}
\end{figure}

\begin{figure}[!h]
\begin{tabular}{cc}
\includegraphics[scale=0.42]{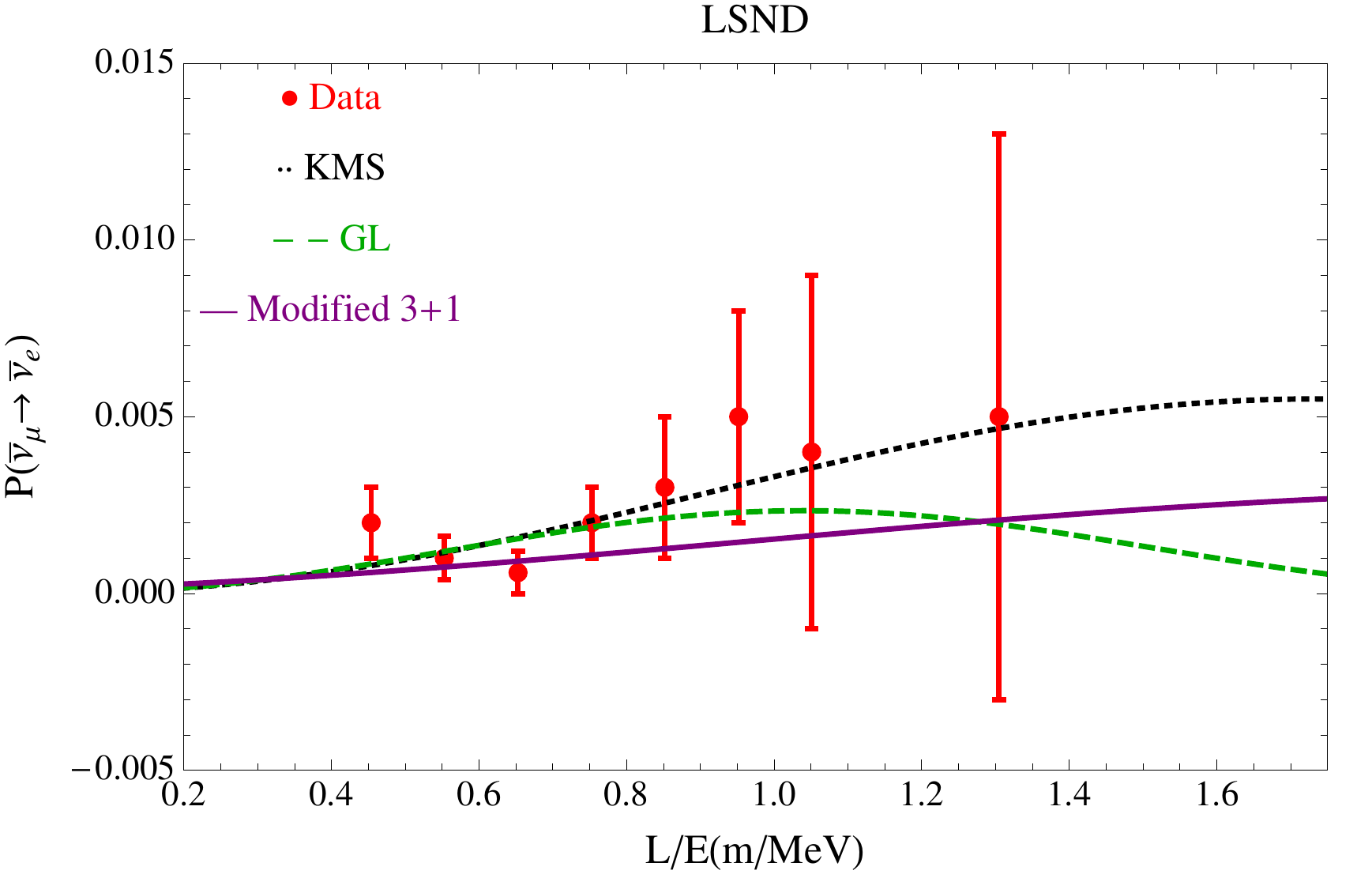}& 
\includegraphics[scale=0.42]{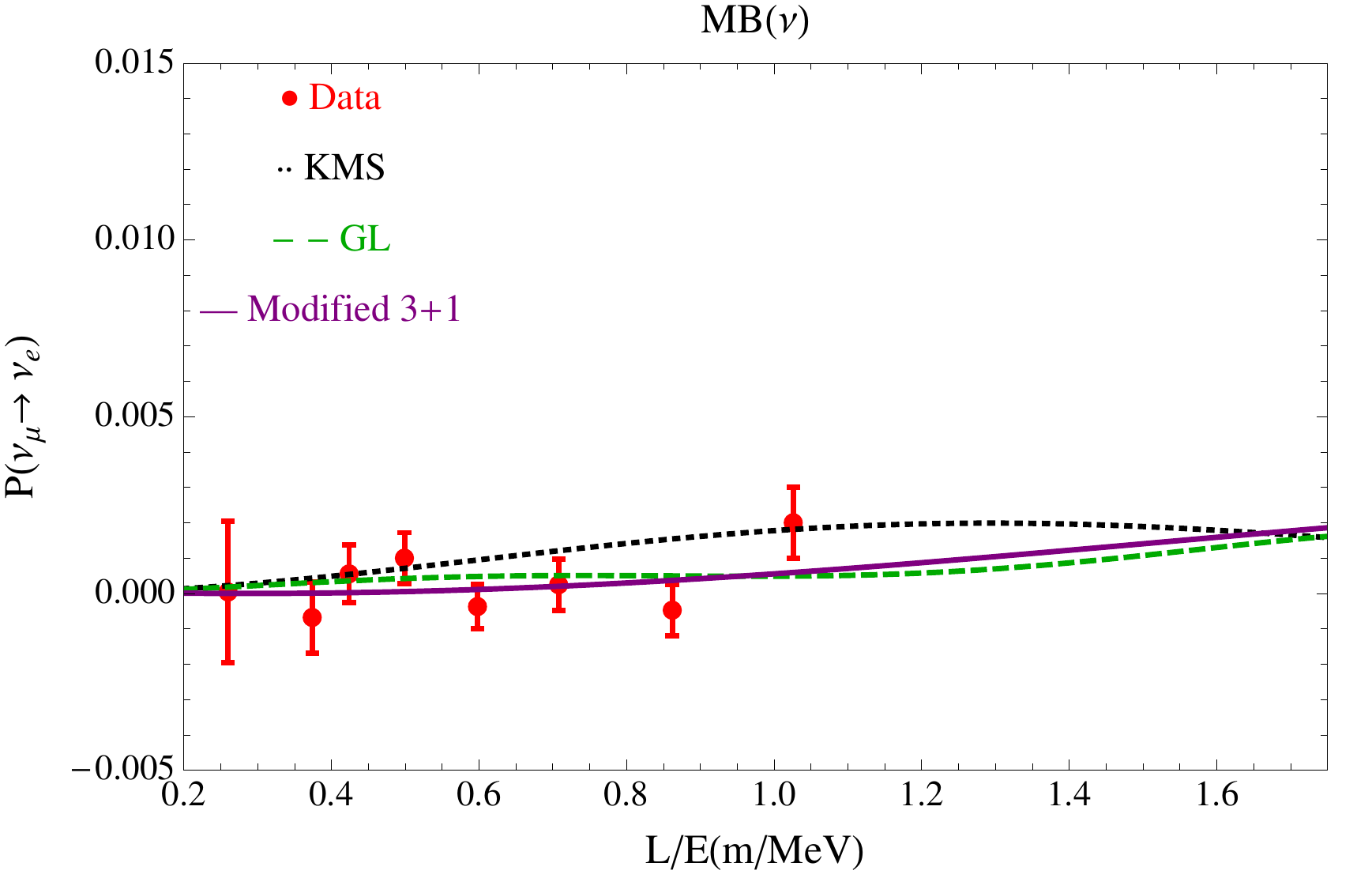} \\
 \includegraphics[scale=0.42]{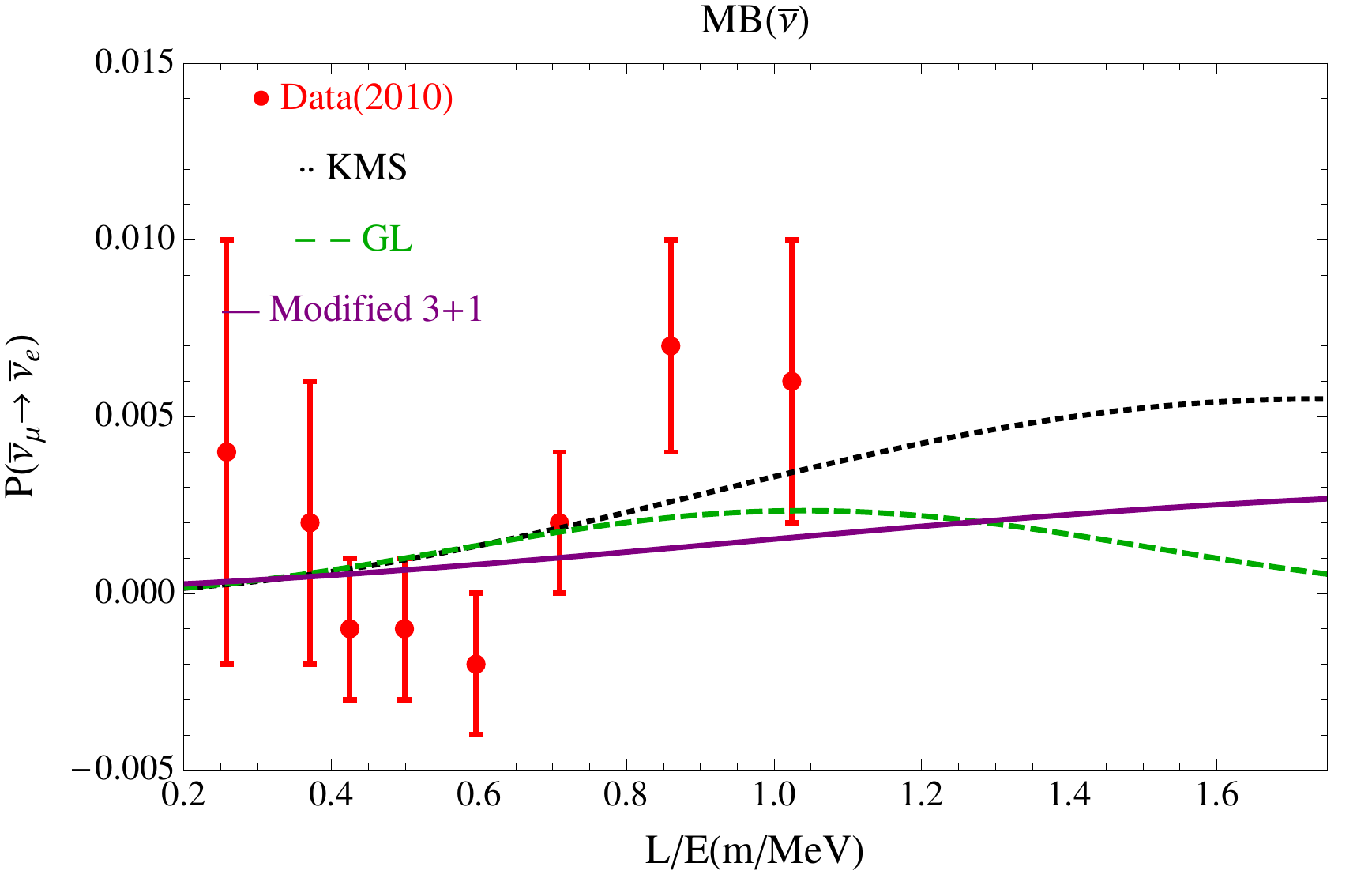}&
  \includegraphics[scale=0.42]{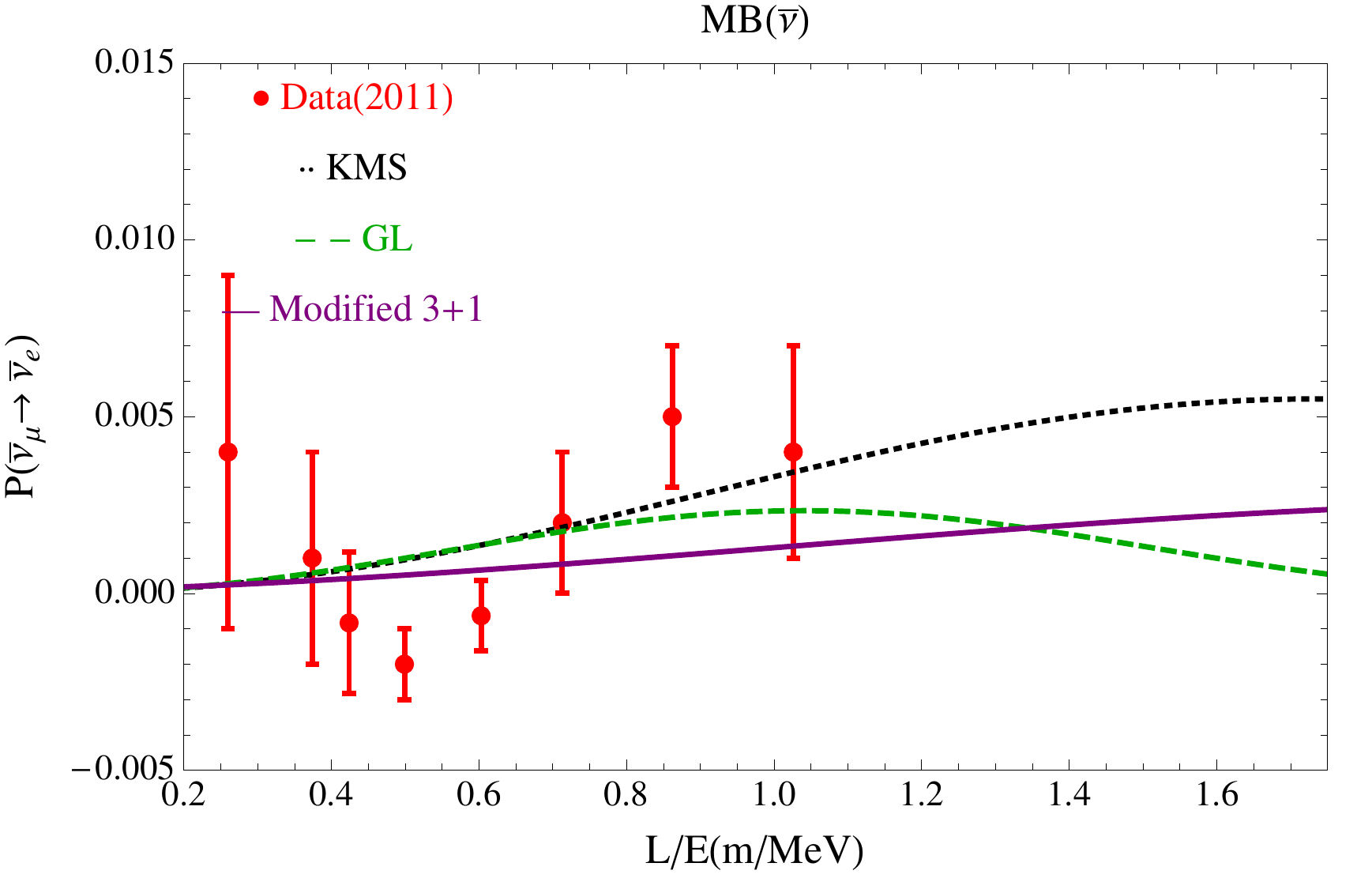}
\end{tabular}
\caption{The predicted probabilities for LSND, MiniBooNE neutrino, and MiniBooNE anti-neutrino oscillations in the
Modified $3+1$ scheme at $\Delta m_{41}^2 = 0.6\,{\rm eV}^2$, $a=0$, and $|U_{e4}U_{\mu4}|, r, \beta$ chosen to minimize $\chi^2$. Also shown for comparison are the global best fit results in the $3+2$ model. }
\label{fig:heavyprob}
\end{figure}

A  $\chi^2$ fit has already been performed by Nelson~\cite{Nelson:2010hz}, which indicates that  the modified $3+1$ scheme with CP violation could possibly account for the short baseline anomalies. Here, we perform a similar fit to allow easy  comparison with the minimal mini-seesaw fits in Section~2 and to incorporate the MiniBooNE 2011 data.  We follow a similar procedure
to~\cite{Nelson:2010hz},  with the probability given by Eq.~\ref{eq:oscillation2}. We numerically evaluated a $\chi^2$ function similar to Eq.~\ref{eq:chisquare} with the free parameters  $|U_{e4}U_{\mu4}|, r, \beta, m_{41}$ while fixing $a$. (The fit does not vary much with different values of $a$.) We include 8 bins each from LSND and MiniBooNE neutrino and anti-neutrino data as  in Section~2.4. Following~\cite{Nelson:2010hz}, we also include  one bin for KARMEN and one for NOMAD~\cite{Astier:2003gs}, with the experimental errors chosen to be the 90\% upper limit on the average oscillation probability. We also impose the 
constraint $|U_{e4}U_{\mu4}| \leq 0.03$ to avoid too large effects in reactor and other experiments.
The total number of fit points included is 26 with 4 free parameters. For the MiniBooNE 2010 data, at $a=0$, the best fit point has $|U_{e4}U_{\mu4}|=0.03, r= 1.10, \beta=-0.15, \Delta m_{41}^2=0.49 \,{\rm eV}^2$ with a $\chi^2=21.5$. For $a=1$, one finds  $|U_{e4}U_{\mu4}|=0.03, r= 1.04, \beta=-0.14, \Delta  m_{41}^2=0.48 \,{\rm eV}^2$ with a $\chi^2=21.9$. The results, which are consistent with those in~\cite{Nelson:2010hz}, are presented in Figure~\ref{fig:chisquare}, which show the $\chi^2$ and $|U_{e4}U_{\mu4}|$ distributions as a function of $\Delta m_{41}^2$, with $|U_{e4}U_{\mu4}|, r, \beta$ chosen to minimize $\chi^2$ at each $\Delta m_{41}^2$. We also show $|U_{e5}U_{\mu5}|$, as calculated from Eq.~\ref{eq:mixing5}. We  caution that, similar to Section~2, this fit does not incorporate all data, except for the imposed upper limits on the mixings, and should only be viewed as evidence for the modified $3+1$ scheme with CP violation as a solution to short baseline neutrino anomalies. The probability distributions for the short baseline experiments at one particular point in the parameter space with a small $\chi^2$ are shown in Figure~\ref{fig:heavyprob}. The quality of the fit gets worse when including 2011 MiniBooNE data. For instance, at $a=0$, the best fit point has $|U_{e4}U_{\mu4}|=0.03, r= 1.08, \beta=-0.12, \Delta m_{41}^2=0.47 \,{\rm eV}^2$ with a $\chi^2=27.5$.

\subsection{Constraints}
One crucial feature of the results in the previous section is that the heavy sterile neutrino, independent of its mass, needs to have large mixing angles with $\nu_e$ and $\nu_\mu$, e.g., $|U_{e5}U_{\mu5}|\gsim 10^{-2}$, to yield large enough CP violation required for the explanations of the short baseline anomalies. Similar to  light sterile neutrinos, a heavy sterile neutrino with large mixing is also constrained by CMB, LSS and BBN data. There are also several other laboratory and cosmological constraints. All of the constraints are summarized in Figures 1 and 2 in~\cite{Smirnov:2006bu, Kusenko:2009up}, to which we refer the reader for more details. Below we just review the most stringent ones in different sterile neutrino mass ranges. 

For a sterile neutrino with mass above 1 eV but below 5 keV, LSS sets the strongest bound on the energy density stored in the sterile neutrino as a function of its mass, and rules out mixing of order $0.1$. The heavy sterile neutrino could decay to an active neutrino and a photon at the one-loop order: $\nu_s \to \nu_a + \gamma$ with $E_\gamma \approx m_S/2$. Thus the X-ray emission from galaxy clusters such as the Virgo cluster or from all possible sources seen as the diffuse extragalactic background impose a strong limit on the flux from the  decay. The bound is strongest in the mass range from a few keV to around a few hundred keV, restricting the mixing-squared to be smaller than at least $10^{-5}$. If the light active neutrinos from the decay are produced between the active neutrino and photon decoupling time, they would contribute to the energy density of the relativistic species and modify the CMB angular power spectrum. The CMB spectrum rules out a heavy sterile neutrino with mass from $\calo$(100 keV - 10 MeV) with mixing $\gsim 10^{-2}$. Sterile neutrinos with mass 1 MeV $\lesssim m_s \lesssim$ 400 MeV are constrained by the non-observation of additional lines in the charged lepton spectrum and the decay products from a heavy sterile neutrino produced by meson decays such as $(\pi^+, K^+) \to \mu^+ \nu_\mu$~\cite{Kusenko:2004qc}, which also rule out mixing of order 0.1. For an even heavier sterile neutrino with mass $\gsim 0.1$ GeV, there are weak constraints on the mixing $\lesssim (0.03 - 0.1)$ depending on the mass from flavor changing neutral currents and violation of lepton universality~\cite{Nardi:1994iv, Atre:2009rg}. In addition, LEP searches for a SM singlet neutral heavy lepton $N_l$ in the channel $Z \to N_l \nu_l$ exclude the mixing-squared down to below $10^{-2}$ in the mass range between 1 GeV and 90 GeV~\cite{Adriani:1992pq}. Thus, without modifying any cosmological assumption, there is only room for a heavy sterile neutrino with mixing $\sim 0.1$ for mass in the range (0.4 - 1) GeV . 

Finally we  comment that there are two additional ways to avoid the cosmic constraints  for heavy sterile neutrinos. The first scenario has a low reheating temperature after inflation, $T_R \ll 100$ MeV to dilute the sterile neutrino density and relax the cosmic bounds, but $T_R > 5$ MeV to preserve a successful BBN~\cite{Gelmini:2004ah}. Another possibility is to have the sterile neutrino decay to a massless or very light (pseudo-)Goldstone boson, e.g., Majoron, with a very short lifetime $\tau \ll 1$ s~\cite{PalomaresRuiz:2005vf}.

\section{Conclusion}
Motivated by short baseline neutrino oscillation anomalies, we study two scenarios with light eV-scale sterile neutrino(s). The first  has two eV-scale  sterile neutrinos and a mass pattern arising naturally from the mini-seesaw mechanism. The mini-seesaw  could be generated by a set of higher-dimensional operators involving a new TeV-scale VEV, which could be related to TeV-scale flavor models. In the limit where the active neutrino Majorana mass vanishes, a simple formula is derived relating the 6 complex mixing parameters between the sterile and active neutrinos to only 5 real parameters. A simple numerical fit is performed based on this minimal mini-seesaw formula, indicating that it could account for the short baseline anomalies. We also review the current laboratory and cosmological constraints and implications for this scenario. The second paradigm has one eV-scale sterile neutrino with a CP-violating phase generated by integrating out a heavy sterile neutrino. However, for large enough CP violation to explain the anomalies, the heavy sterile neutrino has to have a large (fine-tuned) mixing with the active ones, which is in strong tension with astrophysical and laboratory observations. Ongoing and future neutrino oscillation experiments, as well as  cosmological observations such as the {\em Planck} mission and laboratory searches for neutrino mass
effects in $\beta$ decay and neutrinoless double $\beta$ decay will hopefully confirm or exclude the existence of light
sterile neutrinos as an explanation of the short baseline anomalies.

\acknowledgments{We thank Tao Liu for collaboration at an early stage of this project. We also thank Nima Arkani-Hamed for an illuminating discussion. J.F. is supported by the DOE grant DE-FG02-91ER40671.}

\appendix
\section{More details of the Fit with the Minimal Mini-Seesaw Parameterization}

In Section~\ref{sec:miniseesaw fit}, we present a simple fit based on Eq.~\ref{ng1}, using
Mathematica  to calculate the  $\chi^2$ and find the best fit point. 
The result, listed in Table~\ref{fitresults},
 uses data from LSND and MiniBooNE,
as well as a constraint that the active-sterile mixing matrix elements should not exceed 0.15 in magnitude. 
If we had not required $|U_{\alpha i}| < 0.15$,
some of the mixings would have been even larger at the best fit point (the largest was $|U_{e4}| \sim 0.3$). This is a
 reflection of the well-known fact that there is significant tension between the short baseline data and other reactor and accelerator constraints~\cite{Akhmedov:2010vy,Kopp:2011qd,Giunti:2011gz,Maltoni:2007zf,Giunti:2011hn}.
 Nevertheless, the restricted fit has a quite reasonable $\chi^2$, as seen in Table~\ref{fitresults}.
 
\begin{figure}[!h]
\begin{tabular}{cc}
\includegraphics[scale=0.5]{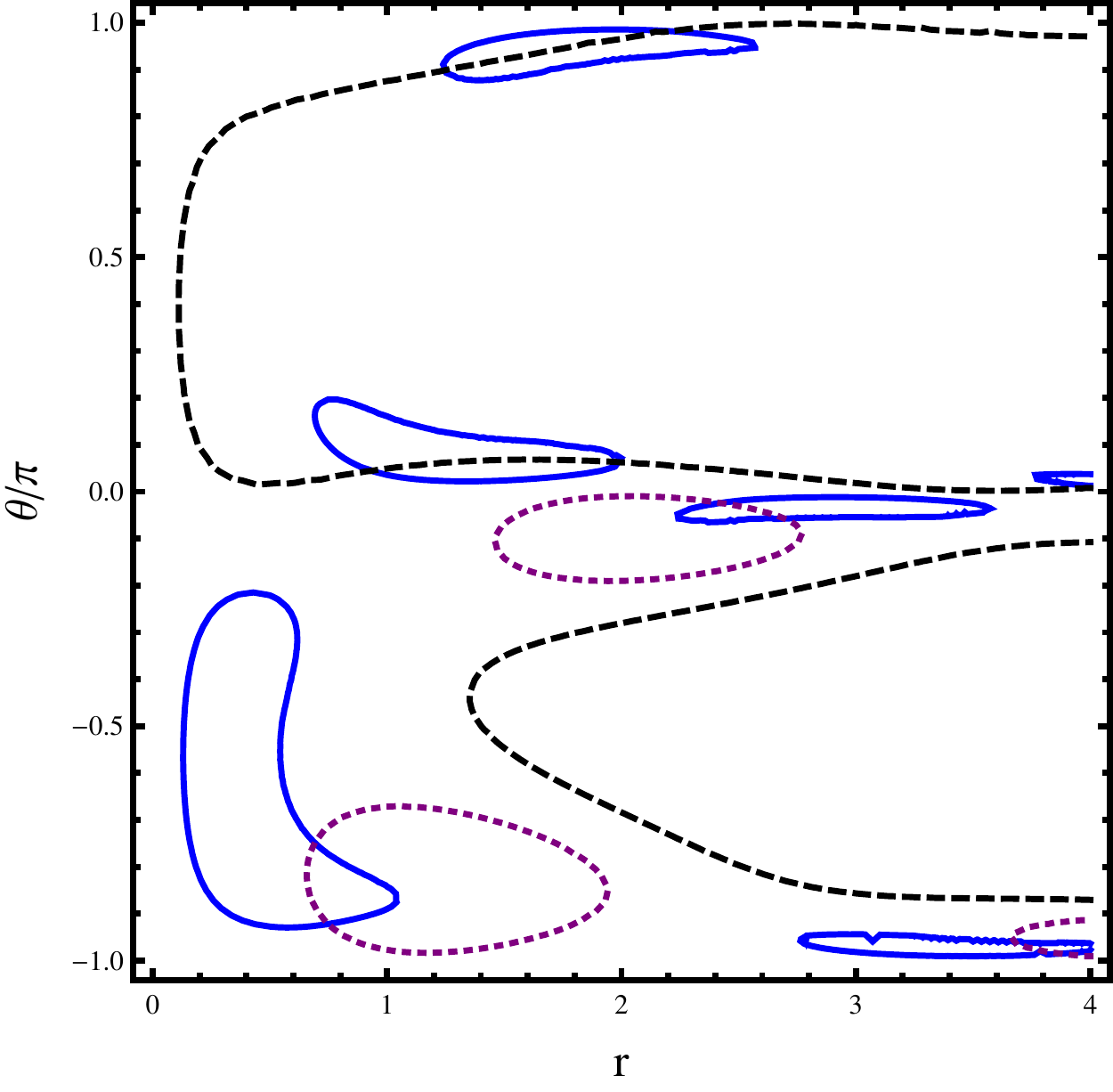}& 
\includegraphics[scale=0.5]{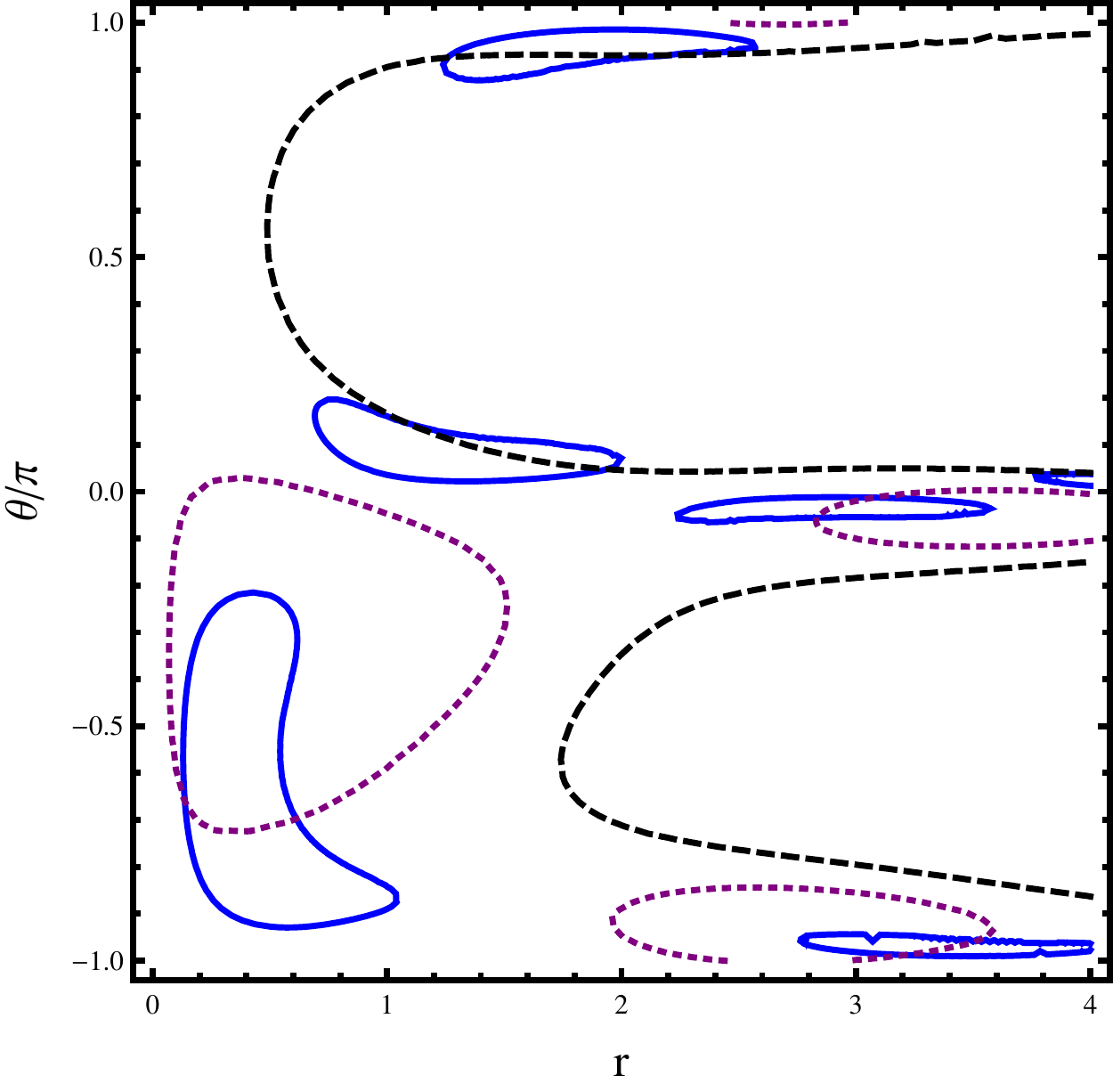} 
\end{tabular}
\caption{Allowed regions in the ($r, \theta$) plane at 95\% C.L (blue, solid) and contours of mixing $|U_{e4}|$ (left) and $|U_{e5}|$ (right) equal to 0.1 (purple, dotted) and 0.2 (black, dashed). }
\label{fig:compare}
\end{figure}

\begin{figure}[!h]
\begin{center}
\includegraphics[scale=0.63]{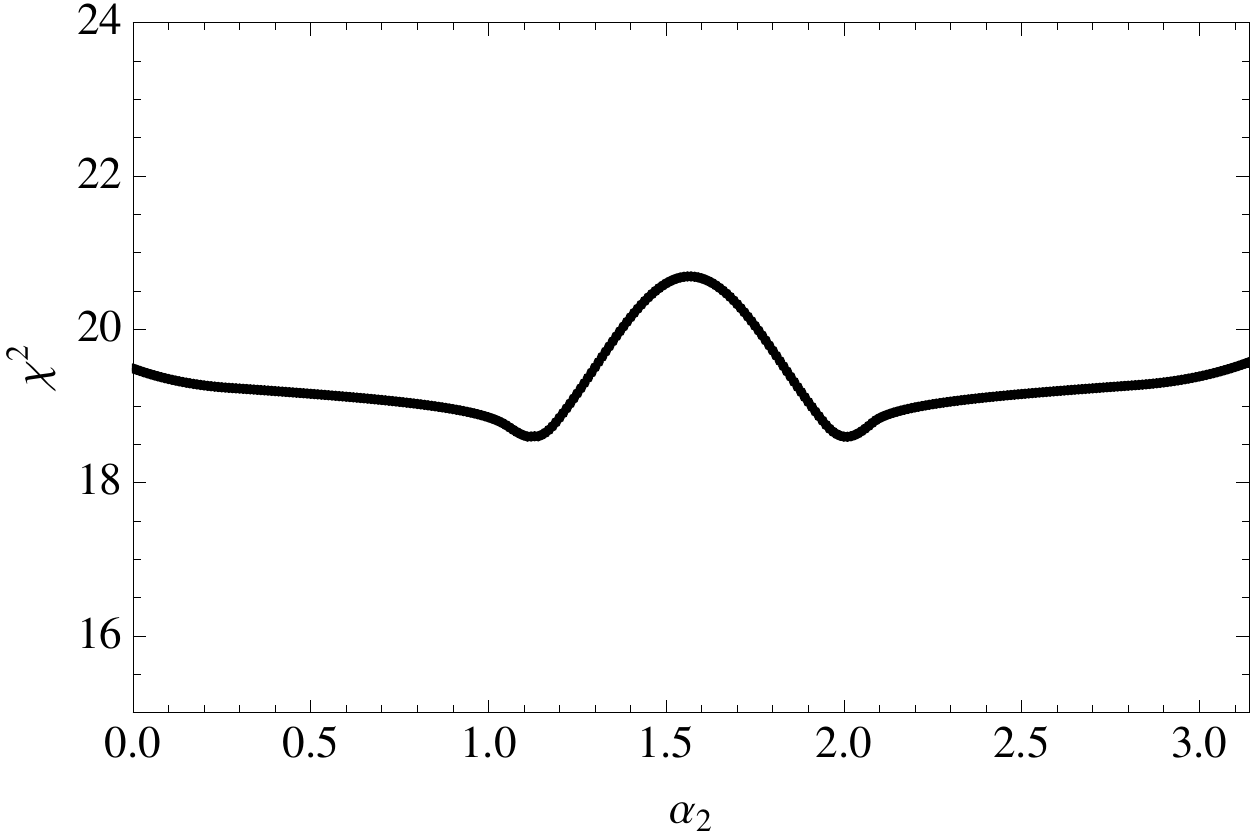}
\end{center}
\caption{The minimal $\chi^2$ as a function of the phase $\alpha_2$ in the active sector with all other parameters $z, M_4, M_5$ chosen to minimize $\chi^2$. The input experimental data is LSND and MiniBooNE 2010 data.}
\label{fig:chisquarealpha2}
\end{figure}

 We compare the contours of mixings and of the allowed region ({\em not} imposing the mixing contraint)  in the ($r, \theta$) plane at 95\% C.L. in Figure~\ref{fig:compare}. It shows that most of the allowed region leads to reasonably small mixings $\lesssim 0.2$. We also show the dependence of $\chi^2$ as a function of the phase $\alpha_2$ in the active sector in Figure~\ref{fig:chisquarealpha2}. The $\chi^2$ does not change much over the entire range, with two minima at $\alpha_2 =1.2, 2$, where the latter has a slightly smaller $\chi^2$.\footnote{This is an accident of the parameters. The distribution would be symmetric under $\alpha_2 \to \pi-\alpha_2$ for $\theta_{13}=0$ and $m_1=m_2$ in the IH.} 

\bibliography{ref}
\bibliographystyle{jhep}

\end{document}